\documentclass[usenatbib]{mn2e}

\usepackage[letterpaper,totalwidth=480pt,totalheight=680pt]{geometry}

\usepackage{amsmath,amssymb} \usepackage{ifthen} \usepackage{comment} 

\newboolean{includeversions}

\newboolean{blackandwhite} 

\newboolean{usepdf} \newif\ifpdf \ifx\pdfoutput\undefined \pdffalse \else \pdfoutput=1 \pdftrue \fi

\ifpdf \setboolean{usepdf}{true} \else \setboolean{usepdf}{false} \fi

\ifthenelse {\boolean{usepdf}}{ \usepackage[pdftex,final]{graphicx} \usepackage[pdftex,draft]{hyperref} }{ \usepackage[dvips,final]{graphicx} \usepackage[dvips]{hyperref} } \graphicspath {{../../../thesis/figures/}{./}} 

\newcommand{\unitvec}[1]{\bmath{\hat #1}}
\renewcommand{\vec}[1]{\bmath{#1}}

\newcommand{\um}{\, {\mu \rm m}}

\newcommand{\kpc}{\, {\rm kpc}}

\newcommand{\Angstrom}{\, {\rm \AA}}

\newcommand{\del}{\nabla}
\newcommand{\grad}{\del}

\newcommand{\identicallyeq}{\equiv}

\newcommand{\Msun}{\,\mathrm{M}_{\odot}}

\newcommand{\pc}{\mathrm{pc}}
\providecommand{\halpha}{{\mathrm{H}\alpha}}
\providecommand{\hbeta}{{\mathrm{H}\beta}}
\renewcommand{\halpha}{{\mathrm{H}\alpha}}
\renewcommand{\hbeta}{{\mathrm{H}\beta}}
\newcommand{\mcrx}{\textit{Sunrise}}
\providecommand{\ga}{\gtrsim}
\providecommand{\la}{\lesssim}
\providecommand{\hii}{\mbox{H\,{\sc ii}}}
 \usepackage {astrojournals} 
\renewcommand{\um}{\, {\umu \rm m}}  \renewcommand{\mcrx}{{\sc sunrise}}

\title {{\sc Sunrise}: Polychromatic Dust Radiative Transfer in   Arbitrary Geometries} \author[Patrik Jonsson] {Patrik Jonsson$^1$\thanks{E-mail address:     patrik@ucolick.org}\\ $^1$Santa Cruz Institute for Particle Physics, University of   California, Santa Cruz, CA 95064} 
\begin {document}

\date{\ifthenelse {\boolean{includeversions}}{  Draft Version: \$Id: sunrise.txt,v 1.41 2006/07/21 22:06:06 patrik Exp $ $ }{} } \pagerange{\pageref{firstpage}--\pageref{lastpage}} \pubyear{2006}

\maketitle

\label{firstpage}

\begin{abstract}
  This paper describes \mcrx , a parallel, free Monte-Carlo code for
the   calculation of radiation transfer through astronomical dust. 
\mcrx \    uses an adaptive-mesh refinement grid to describe arbitrary 
 geometries of emitting and absorbing/scattering media, with spatial  
dynamical range exceeding $10^4$, and it can efficiently   generate
images of the emerging radiation at arbitrary points in   space.  In
addition to the monochromatic radiative transfer   typically used by
Monte-Carlo codes, \mcrx \ is capable of propagating   a range of
wavelengths simultaneously.  This ``polychromatic''   algorithm gives
significant improvements in efficiency and   accuracy when spectral
features are calculated. \mcrx \ is used to   study the effects of dust
in hydrodynamic simulations of interacting   galaxies, and the
procedure for this is described.  The code is   tested against
previously published results.
\end{abstract}



\section{Introduction}

Interstellar dust profoundly affects our view of the universe, from
obscuring the stars forming in giant molecular clouds in our Galaxy, to
camouflaging extreme starbursts as relatively unremarkable galaxies in
Ultraluminous Infrared Galaxies, to completely hiding high-redshift
galaxies from our view except in the infrared, as in the sources
detected with SCUBA.  Because of this, modeling the effects of dust has
been the subject of an ever-increasing number of papers.  Initial
models used very simple assumptions, such as the dust being distributed
as a foreground screen.  While appropriate for observations of single
stars, this assumption fails miserably in the case of galaxies, where
the stars and dust are intermixed.  In this situation, the appearance
of the system will depend not only on the characteristics of the dust
itself but also on the relative distributions of stars and dust.  In
this scenario analytic solutions in general do not exist, and numerical
solutions become necessary.

Numerical approaches to solving the radiative-transfer problem for dust
can generally be classified as either finite-difference methods
\citep[e.g.][]{steinackeretal03, folinietal03},
or Monte-Carlo methods, where the problem is solved in a stochastic
sense.  The advantages of Monte-Carlo methods are that they can easily
treat complications such as arbitrary geometries and nonisotropic
scattering, the main drawback being that generating a sufficiently
accurate solution can be computationally quite expensive.
Traditionally, the computational expense has been reduced by using
simple geometries and exploiting symmetries in the problem, such as
assuming spherical or azimuthal symmetry.  It is only more recently
that fully three-dimensional, arbitrary geometries have been tackled
\citep{wolfetal99, bianchietal00, gordonetal01, kurosawahillier01,
baesetal03, harriesetal04}. Codes combining dust and photoionization
radiative transfer are also appearing \citep{ercolanoetal05}.

This work describes \mcrx , a new code for Monte-Carlo radiative
transfer calculations.  The development of this code was motivated by a
desire to calculate the effects of dust in a large program of
hydrodynamic N-body simulations of interacting galaxies
\citep{tjthesis, coxetal05methods}.  The main requirements for this are
that the code be able to handle arbitrary geometries and resolve the
large spatial dynamic range in the simulations, and that the large
number of calculations necessary can be completed in reasonable time. 
These requirements made using an adaptive-mesh grid to represent the
simulation geometry a necessity, and the code was parallelized to
handle the demanding computational requirements.  Another desired goal
was the ability to efficiently predict spectral features, such as
emission lines, the $4000 \Angstrom$ break and the UV continuum slope,
that are used in observational studies of galaxies. The
``polychromatic'' implementation of the Monte Carlo method presented
here achieves this goal.  While hydrodynamic simulations would seem to
be an ideal source for geometry-dependent radiative-transfer problems,
this has not been done more than occasionally.  Those cases have either
concerned protostars   \citep{fischeretal94} or star-forming regions
\citep{kurosawaetal04}, or, if they have simulated galaxies, have used
more schematic treatment of the hydrodynamics and the radiative
transfer \citep{bekkishioya00a, bekkishioya00b, cattaneoetal05}.  To
our knowledge, this work is the first that combines SPH hydrodynamic
simulations including supernova feedback with a full radiative-transfer
model to study the effects of dust in galaxies.  Results from the
simulations have been presented in \citet{pjthesis} and
\citet{pjetal05attn}, and additional papers are in preparation.

While a number of implementations of Monte-Carlo radiative transfer
codes are described in the literature, none of these are publicly
available.  This is in marked contrast to hydrodynamic codes, of which
several are publicly available.  As a service to the community, the
author is releasing \mcrx \ as free software.

The organization of this paper is as follows:  in
section~\ref{background_section}, an overview of the radiative-transfer
problem and the Monte-Carlo method is given. 
Section~\ref{radiative_transfer_section} describes the core
radiative-transfer algorithm, while
section~\ref{auxiliary_calculations_section} describes the additional
steps necessary to apply the radiative-transfer calculations to the
output of hydrodynamic simulations.    The code is tested against
previously published results in section~\ref{test_section}. In
section~\ref{implementation_section}, implementation details are given
and finally prospects for future improvements are given in
section~\ref{future_section}.

\section{Background}
 \label{background_section}

\subsection{The Radiative-Transfer Problem }
 \label{rte_section}

The radiative-transfer problem is the problem of calculating the
propagation of radiation through a medium which may emit, absorb or
scatter the radiation. In the case of the problem of the transfer of
stellar radiation through astrophysical dust, there are a number of
simplifications that can be made, which greatly reduces the complexity
of the problem. First, astronomical systems  evolve slowly enough that
the time dependence  practically always can be ignored, and the
equation to be solved is the time-independent radiative-transfer
equation: 
\begin{equation}
\label{equation-tid-rte} \unitvec k \cdot \vec \grad I_\nu + \rho
\kappa_\nu I_\nu = \rho \left ( \frac { j_\nu } { 4 \pi } +
\kappa_\nu^{ \rm { sca } }
\int \phi_\nu ( \unitvec k , \unitvec { k ' } ) I_\nu ( \unitvec { k '
} ) \mathrm {d} \Omega '
\right ) \>.
\end{equation}
 The dependent variable, $I_\nu$, is the ``specific intensity'',
defined by 
\begin{equation}
\mathrm {d} E = I_\nu ( \unitvec k , \vec x , t ) \unitvec k \cdot
\bmath { \mathrm {d} } \vec A \, \mathrm {d} \Omega \, \mathrm {d} \nu
\, \mathrm {d} t \>.
\end{equation}
 In other words, it is the amount of energy per unit time per area per
solid angle per frequency interval flowing in direction $\unitvec k$.
$\kappa_\nu$ is the total opacity of the medium. The terms on the
right-hand side of Equation~\ref{equation-tid-rte} represent emission
of radiation and the addition to the intensity from radiation scattered
into the direction $\unitvec k$ from other directions.

The emission from stars is fixed and does not depend on the local
radiation field. There is emission from dust grains, thermal emission
by the grains which are heated by the radiation field, but the dust
grains emit mostly at wavelengths where stars do not and where the dust
itself is largely optically thin. This means that the contribution from
dust emission at wavelengths where stellar emission is important can be
ignored except for a small wavelength interval around $5 \um$ and that
the contribution to the heating of dust grains by the emission from
other dust grains is a second-order effect only important in very
optically thick regions.

Furthermore, we do not require knowledge of the full radiation field at
all points in space; normally it is sufficient to know the radiation
field that escapes to infinity, which is what reveals the external
appearance of the object, and the average intensity at points inside
the object, which is what determines the heating of the dust grains. 
It is the presence of the scattering integral in
Equation~\ref{equation-tid-rte} which poses the largest difficulty. 
Scattering, however, is easily accounted for when using Monte Carlo
methods.

\subsection{The Monte-Carlo Method}
 \label{monte_carlo_section}

The Monte-Carlo method is a way of solving equations by random
sampling, and its usefulness for solving particle transport problems
was first recognized by Fermi, Ulam, and von Neumann during the days of
the Manhattan Project.  For an overview of the Monte-Carlo method as it
pertains to particle transport problems, see e.g. \citet{mcprimer} or,
for more advanced topics, \citet{luxkoblinger91}.  Essentially, Monte
Carlo is the ``natural'' way of solving the radiative transfer
equation, in the sense that the photons in nature are unaware that they
are solving a very difficult equation, they are simply obeying the
local conditions.  In the same way, solving the RTE using the
Monte-Carlo method amounts to statistically sampling the processes that
emit, scatter and absorb photons.  After tracing many such photons,
statistics build up and the solution emerges.  This local
characteristic also makes the Monte-Carlo method naturally able to
handle arbitrary geometries and complicated scattering characteristics

\subsection{Drawing Random Numbers}
 \label{section_drawing}

Random numbers are at the heart of the Monte-Carlo method.  The ability
to draw random numbers with various probability distributions is
essential.

Given a Probability Density Function (PDF) $f ( x )$ for a stochastic
variable X, if we generate a random number $R$, uniformly distributed
on $[ 0 , 1 ]$, solving the equation 
\begin{equation}
\label{equation-draw-variable} \int\limits_l^x f ( x ' ) \,\mathrm {d}
x ' = R
\end{equation}
 yields a number $x$ with a PDF $f ( x )$. ($R$ will be used throughout
this paper to denote \emph{a realization} of a stochastic variable
distributed uniformly over $[ 0 , 1 ]$.  This means that $R$ in one
expression is never equal to $R$ in another, just that they were drawn
from the same PDF.)

As an example, let us derive the PDF of the optical depth at which a
photon will interact.  Photons propagating through an opaque medium
interact after traversing different optical depths --- most quickly,
before traversing unit optical depth, while a few make it through many.
  These optical depths are distributed as 
\begin{equation}
P ( \tau ) = e^{ - \tau } \>.
\end{equation}
 The optical depth $\tau_i$ at which a photon will interact with a
medium can then be randomly generated using
Equation~\ref{equation-draw-variable}: 
\begin{eqnarray}
\label{equation-draw-optical-depth} \int\limits_0^{ \tau_i } e^{ - \tau
' } \,\mathrm {d} \tau ' = R & \Rightarrow & \tau_i = - \ln ( 1 - R )
\identicallyeq - \ln R \>.
\end{eqnarray}
 (The last equivalence in Equation~\ref{equation-draw-optical-depth} 
may look bizarre until it is realized that $1 - R$ has the same PDF as
$R$!)  Equation~\ref{equation-draw-optical-depth} is the formula used
to randomly draw interaction optical depths in the code.

\subsection{Biasing}
 \label{biasing_section}

It should be noted that Equation~\ref{equation-draw-optical-depth} is
not a unique (and, depending on the situation, not even necessarily the
best) way of sampling interaction optical depths.  The probability
distribution from which a quantity is drawn can be ``biased'' to sample
certain parts of the distribution more heavily.  In order to preserve a
statistically correct result, the difference in probability must be
compensated for by assigning a ``weight'' to the samples.  In general,
if a process with the probability distribution $f ( x )$ is sampled
with another probability distribution $g ( x )$, the samples must be
weighted by the factor $w = f ( x ) / g ( x )$.  The only requirement
on $g ( x )$ is that $g ( x ) > 0$ for all $x$ for which $f ( x ) >
0$.

Biasing can be a very effective way of reducing the variance in the
results of a Monte Carlo calculation by more effectively sampling the
parts of the distribution that are important for the end result
\citep{mcprimer}.  \citet{juvela05} explored different biasing schemes
in the context of radiative transfer through a spherically symmetric
dust cloud, and found potential increases in efficiency by more than an
order of magnitude.  However, selecting a proper biasing requires a
priori knowledge of the specific problem. Biasing is also the
theoretical basis for the polychromatic algorithm described in
Section~\ref{polychromatic_section}, since it allows different
wavelengths, with different mean free paths, to be sampled with
identical rays.

\section{The Radiative-Transfer Algorithm}
 \label{radiative_transfer_section}

As explained previously, the radiative-transfer problem will be solved
through statistical sampling of the processes of photon emission,
scattering and absorption.

In \mcrx , the dust opacity is represented on an adaptive grid, the
characteristics of which are described in
Section~\ref{adaptive_grid_section}.  There are many possible sources
of emission, for example collections of point sources, external diffuse
radiation or emission continuously distributed on the same adaptive
grid.  In the galaxy merger simulations, the emission comes from the
finite-sized ``stellar particles'' tracked by the SPH code.  The use of
an adaptive grid allows the representation of arbitrary geometries,
limited only by the amount of computer time dedicated to running the
problem.  (In principle, memory is also a limitation, but in practice
it has been found that memory use is much less of a constraint.)

\subsection{Ray Tracing}

\begin{table*}   \begin {center} \begin{tabular}{ll} Symbol & Description \\ \hline $\kappa$ & Dust opacity, or interaction cross-section per unit mass. \\ $a$ & Dust grain albedo (fraction that is scattered during an interaction). \\ $g$ & Scattering phase function asymmetry, $\left<\cos\theta\right>$. \\ $\mathcal{L}_e$ & Luminosity of emitter $e $. \\ $I_{i,j}$ & Intensity (normally $\in [0,1]$) of the $i$'th ray after the $j$'th interaction. \\ & \quad($j=0$ is the point of emission.) \\ $L_\mathrm{tot}$ & Total luminosity of the system. \\ $N_\mathrm{tot}$ & Total number of rays traced. \\ $n$ & Luminosity normalization, $L_\mathrm{tot}/N_\mathrm{tot}$. \\ $h$ & SPH smoothing length of the particles \\ $V_c$ & Volume of cell $c$. \\ $\Delta\ell_{i,j,c}$ & Path length for which the $i$'th ray, after the $j$'th interaction, passes through \\ & \quad cell $c$. \\ $L_i$ & Luminosity associated with the $i$'th ray. \\ $F_{i,j}$ & Flux reaching the observer from the $j$'th interaction site of the $i$'th ray. \\ $\unitvec{k_{i,j}}$ & Direction vector of the $i$'th ray after the $j$'th interaction. \\ $\unitvec{k_{i,j}^{\rm{obs}}}$ & Direction vector towards the observer from the $j$'th interaction site of the \\ & \quad $i$'th ray. \\ $d_{i,j}$ & Distance from the the $j$'th interaction site of the $i$'th ray to the observer. \\ $\tau_{i,j}^{\rm{obs}}$ & Optical depth from the the $j$'th interaction site of the $i$'th ray to the observer. \\ $\tau_{i,j}^{\rm{e}}$ & Optical depth from the the $j$'th interaction site of the $i$'th ray \\&\quad to the edge of the medium. \\ $\tau_{i,j}$ & Randomly drawn interaction optical depth of the $j$'th interaction of the\\&\quad $i$'th ray. \\ $\Phi_e(\mathbf{\hat r})$ & Angular distribution of emitted radiation. \\ $\Phi_s(\mathbf{\hat r},\mathbf{\hat r'})$ & Probability distribution of radiation from direction $\mathbf{\hat r}$ scattering into \\&\quad direction $\mathbf{\hat r'}$. \\ $\Phi_s(\cos \theta)$ & Probability of radiation scattering an angle $\theta=\mathbf{\hat r} \cdot \mathbf{\hat r'}$ (scattering phase \\&\quad function). \\ $A_c$ & Absorbed luminosity in cell $c$. \\ $J_c$ & Radiation intensity in cell $c$. \\ $\Sigma_p$ & Surface brightness in pixel $p$. \\  \hline \end{tabular} \end {center} \caption[Symbols used in the description of the radiative-transfer algorithm]{ \label{symbol_table} Symbols used in the description of the radiative-transfer algorithm and their meaning.  There is an implicit dependence on wavelength in all these quantities. } \end{table*}

The simplest implementation of the Monte-Carlo radiative-transfer
algorithm follows a single photon through the medium.  This photon is
emitted in a random direction and can then scatter and/or be absorbed.
Eventually the photon leaves the medium in some direction, which in
general is not the direction from which the object is being imaged.
This method is in general very inefficient.  The efficiency can be
greatly increased by calculating some probabilities analytically by
having each ray represent a (large) number of photons, a ``photon
packet'', whose intensity is proportional to the luminosity in the
packet.  This makes possible the use of inherently probabilistic
constructs like the dust grain single-scattering albedo (the ratio of
the scattering to the total cross-section) to determine the intensity
of scattered radiation, rather than explicitly Monte-Carlo sampling the
absorption and scattering processes, which is much less efficient.  In
general, analytic calculations are more efficient than explicit
Monte-Carlo realization and should be used whenever possible.

Furthermore, since the main object is to generate images of emerging
radiation, it is too inefficient to simply let rays emerge in random
directions.  Instead, an algorithm which efficiently estimates the
radiation which would emerge in the directions from which the system is
imaged is needed.  \mcrx \ uses the ``Next-Event Estimator''
\citep{mcprimer}, which efficiently calculates the flux at a point.
(This method is also described in e.g. \citealt{yusefzadeh84}.)  Using
this estimator, the contribution to the radiation field at the observer
is calculated analytically at each point of emission and scattering. 
The algorithm is described in the following section, using a formalism
similar to that of \citet{gordonetal01}.  For an explanation of the
symbols used, see Table~\ref{symbol_table}.

For the ray tracing, every wavelength is treated independently. This
approach is valid since scattering by dust grains is an elastic
process; the photon is either completely absorbed or scattered without
changing its wavelength.  Most Monte-Carlo codes either trace rays at a
set of discrete wavelengths, or the wavelengths of the rays are sampled
randomly from an appropriate probability distribution.  \mcrx \  has
until now used discrete wavelengths, but here a new approach, where
biasing is used to implement a ``polychromatic'' algorithm where every
ray samples every wavelength, is presented.  The description that
follows initially assumes that the ray tracing is done for one specific
wavelength. The additions necessary for a polychromatic algorithm are
then described in Section~\ref{polychromatic_section}.

\subsection{Dust Properties}
 \label{section-dust-properties}

In order to perform the ray tracing, the properties of the opaque
medium must be specified.  The necessary quantities are $\kappa$, the
mass opacity of the dust grains; $a$, the single-scattering albedo; and
finally the scattering phase function, the angular distribution of the
scattered photons.  While \mcrx \ is capable of using an arbitrary
phase function, the one currently used is the popular (albeit of
questionable accuracy, as pointed out by e.g. \citealt{draine03}) 
\citet[HG,][]{hg41}
functional form, 
\begin{equation}
\label{hg-phase-function} \Phi_s ( \cos \theta ) = \frac { 1 - g^2 } {
4 \pi ( 1 + g^2 - 2 g \cos \theta )^{3/2} } ,
\end{equation}
 where $\theta$ is the scattering angle and $g = \left < \cos \theta
\right >$, the phase function asymmetry, parameterizes the degree to
which the scattering is isotropic or mostly forward/backward.  The HG
function has the advantage that it can be analytically inverted.

In order to calculate the absorption and scattering of light by the
dust, all that is needed are the three quantities $\kappa$, $a$, and
$g$, as a function of wavelength.  Knowledge about the detailed grain
size distribution and composition giving rise to these quantities is
not necessary.  However, if the spectrum of infrared radiation emitted
by the dust is to be calculated, the situation is different. When this
capability, which is a planned upgrade, is added to \mcrx , these
details will be necessary.

\subsection{Emission}

The luminosity associated with ray $i$ at emission is 
\begin{equation}
\label{equation-normalization} L_i = I_{ i , 0 } n \> ,
\end{equation}
 where $I_{ i , j }$ is the intensity of the ray, identifying its
statistical weight, and $n = L_{ \rm { tot } } / N_{ \rm { tot } }$ is
an overall normalization (common for all rays). When rays are emitted,
the probability of emission at a certain point and in a certain
direction is normally equal to the actual probability of emission of
photons, and their initial intensity is unity.  (As explained in
Section~\ref{biasing_section}, this is merely a choice.)

The reason for the separate normalizing factor $n$ is that it is often
desirable to be able to delay the final normalization to until after
all rays have been traced, while $I_i$ must be known at the time the
ray is created.  This way, the total number of rays $N_{ \rm { tot } }$
does not have to be known in advance, which can be the case for example
if the ray tracing is being done on several CPUs in parallel.

In the case of several sources of emission, like a collection of
particles or many grid cells, the emitter $e$ from which the ray is
emitted is drawn randomly by finding $e$ such that 
\begin{equation}
{ \sum_{ e ' = 0 }^e P_{ e ' } } < R \le { \sum_{ e ' = 0 }^{ { e + 1 }
} P_{ e ' } } \> ,
\end{equation}
 where $P_e = L_e / L_{ \rm { tot } }$ is the probability of emission
from emitter $e$.

Once the source of emission has been determined, the originating
position $\vec { x_e }$ is determined.  If the emitter is a
finite-sized particle from the SPH simulations, the radius of the point
of emission is determined based on the density profile represented by
the SPH smoothing kernel.  In order to avoid resorting to numerically
solving for the radius, a polynomial approximation of the probability
distribution resulting from the SPH smoothing kernel is used.  The
simplest polynomial representation that has a mass which goes to 0 at
small radii and a density which goes to 0 at twice the smoothing length
$h$ is 
\begin{equation}
P ( r < ah ) = - a^2 ( a - 3 ) / 4 \> , \> 0 \le a \le 2 \> ,
\end{equation}
 which corresponds to the density distribution 
\begin{equation}
\rho ( a ) = \frac { 3 } { 4 \pi } ( \frac { 2 } { a } - 1 ) \>.
\end{equation}
 The radius of emission is then determined by solving the cubic
equation $P ( r / h ) = R$.

If the source of emission is a grid cell, the position within the cell
is drawn from a uniform random distribution: 
\begin{equation}
\vec { x_e } = \vec { x_{ \rm { min } , c } } + \left ( \vec { x_{ \rm
{ max } , c } } - \vec { x_{ \rm { min } , c } } \right )
\left ( R \unitvec x + R \unitvec y + R \unitvec z \right )
\end{equation}
 where $\vec { x_{ \rm { min } , c } }$ and $\vec { x_{ \rm { max } , c
} }$ are the lower and upper boundaries of the cell. (Remember that the
three instances of $R$ denote three \emph{different} random numbers.)

   With the point of emission determined, the direct flux that would
result from the emission, if the ray was emitted in the direction of
the observer, is calculated: 
\begin{equation}
\label{equation-direct-flux} F_{ i , 0 } = n I_{ i , 0 } e^{ - \tau_{ i
, 0 }^{ \rm { obs } } } \Phi_e ( \unitvec { k_{ i , 0 }^{ \rm { obs } }
} ) \frac { 1 } { d_{ i , 0 }^2 }.
\end{equation}
 $\tau_{ i , 0 }^{ \rm { obs } }$ is the optical depth between the
point of emission and the observer, $\Phi_e$ is the angular
distribution of emitted radiation and $\unitvec { k_{ i , 0 }^{ \rm {
obs } } }$ the direction vector from the site of emission to the
observer.  (In the case of isotropic emission, $\Phi = 1 / ( 4 \pi )$.)
 $d_{ i , 0 }$ is the distance from the site of emission to the
observer. This calculation is repeated to calculate the contribution in
all ``cameras''.

Finally, a specific direction of propagation $\unitvec { k_{ i , 0 } }$
for the ray is randomly drawn from the angular distribution of the
emitted radiation, $\Phi_e$, defined as 
\begin{align}
\Phi_e & = \frac { \mathrm {d} I } { \mathrm {d} \Omega } \> { \mbox {
, such that } } \\ \int \Phi_e \,\mathrm {d} \Omega & = \int \frac {
\mathrm {d} I } { \mathrm {d} \Omega } \,\mathrm {d} \Omega = 1.
\end{align}
 Using Equation~\ref{equation-draw-variable},  it is possible to draw
random directions from this distribution. In the case of isotropic
emission, the procedure is  
\begin{align}
\label{equation-emission-direction} \theta & = \arccos ( 2 R - 1 ) \> ,
\nonumber \\ \phi & = 2 \pi R \> , \\ \unitvec { k_{ i , 0 } } & = (
\sin \theta \cos \phi , \sin \theta \sin \phi , \cos \theta )
\>.\nonumber
\end{align}
 The code also includes other types of emitters, such as point sources,
and other angular distributions of the emitted rays, like collimated
beams.  Arbitrary angular distributions are easily added.

\subsection{Ray Propagation}

The ray is now propagated in the direction $\unitvec { k_{ i , 0 } }$. 
The propagation is done from one cell to another, keeping track of the
optical depth traversed by the ray.  At each step, the optical depth is
increased by $\Delta \tau = \rho_c \kappa \Delta \ell$, where $\rho_c$
is the density of dust in the cell, $\kappa$ is the opacity of the
dust, and $\Delta \ell$ is the path length traversed by the ray inside
the cell.

If the medium traversed is optically thin, most rays would leave the
simulation medium without interacting and not contribute to the
scattered flux.  To increase the efficiency of the calculation of the
scattered flux, \mcrx , like most other Monte-Carlo codes, uses the
concept of ``forced scattering'' \citep{cashwelleverett}, in which
every ray is forced to contribute to the scattered flux. In the
``forced scattering'' scenario, the total optical depth $\tau_{ i , 0
}^e$ from the point of emission $\vec { x_e }$ to the edge of the
medium in the direction of propagation is first calculated (by tracing
the ray to the edge of the grid).  The ray is then split up into two
parts.  One part, $I_{ i , 0 } e^{ - \tau_{ i , 0 }^{ \rm { e } } }$,
will leave the medium without interaction, while $I_{ i , 0 } \left ( 1
- e^{ - \tau_{ i , 0 }^e } \right )$ will interact \emph{somewhere}
along the path.  The optical depth of this interaction, which is in the
range $[ 0 , \tau_{ i , 0 }^{ \rm { e } } ]$, is drawn randomly using
the formula 
\begin{equation}
\tau_{ i , 0 } = - \ln \left [ 1 - R \left ( 1 - e^{ - \tau_{ i , 0 }^e
} \right ) \right ] \> ,
\end{equation}
 which is a variant of Equation~\ref{equation-draw-optical-depth}
obtained by restricting the range of optical depths to $[ 0 , \tau_{ i
, 0 }^{ \rm { e } } ]$ and renormalizing the distribution.  The part of
the ray that leaves the medium is dropped, as the flux resulting from
direct radiation already has been taken into account with
Equation~\ref{equation-direct-flux}.  The part of the ray that does
interact will have an intensity after the interaction of 
\begin{equation}
I_{ i , 1 } = I_{ i , 0 } a \left ( 1 - e^{ - \tau_{ i , 0 }^e } \right
) ,
\end{equation}
 where $a$ is the dust grain albedo.  The luminosity absorbed in the
grid cell where the interaction takes place is 
\begin{equation}
A_{ i , 1 } = n I_{ i , 0 } ( 1 - a ) \left ( 1 - e^{ - \tau_{ i , 0
}^e } \right ).
\end{equation}
 The part of the ray left after the interaction is scattered into a new
direction by the dust grain.  Analogously to
Equation~\ref{equation-direct-flux}, the flux resulting from the part
of the ray which would be scattered towards the observer and which
would not interact on its way there, is 
\begin{equation}
\label{equation-scattered-flux} F_{ i , 1 } = n I_{ i , 1 } e^{ -
\tau_{ i , 1 }^{ \rm { obs } } } \Phi_s ( \unitvec { k_{ i , 0 } } ,
\unitvec { k_{ i , 1 }^{ \rm { obs } } } ) \frac { 1 } { d_{ i , 1 }^2
} ,
\end{equation}
 where $\Phi_s ( \unitvec k , \unitvec { k ' } )$ is the scattering
phase function, i.e. the angular distribution $\mathrm {d} I / \mathrm
{d} \Omega$ for scattering of rays from direction $\unitvec k$ into
direction $\unitvec k '$. In most cases, the phase function will depend
only on the angle between the two directions (exceptions to this would
be e.g. non-spherical dust grains aligned by a magnetic field), in
which case $\Phi_s ( \unitvec k , \unitvec { k ' } ) = \Phi_s (
\unitvec k \cdot \unitvec { k ' } )$.  As mentioned in
Section~\ref{section-dust-properties}, \mcrx \ currently uses the
Henyey-Greenstein function, but is capable of using arbitrary phase
functions.  (Note that the equation equivalent
to~\ref{equation-scattered-flux} in \citealt{gordonetal01}, their
equation~6, has an extraneous $4 \pi$ [K. Gordon 2004, private
communication].)

After calculating the intensity that would have made it to the
observer, had the ray been scattered in that direction, a random
scattering angle is drawn from the scattering phase function.  In the
case of the HG phase function, the formula is \citep{witt77} 
\begin{eqnarray}
\label{hg-draw-theta} \cos \theta & = & \frac { 1 } { 2 g } \left [ 1 +
g^2 - \left ( \frac { 1 - g^2 } { 1 - g + 2 g R } \right )^2 \right ]
\> , \\ \phi & = & 2 \pi R \>.
\end{eqnarray}
 (The arbitrary reference axis for the azimuthal angle $\phi$ is taken
to be the $\unitvec z$ axis.)  The ray is then rotated by this angle,
and this becomes the new direction of propagation.

Depending on the problem at hand, the number of scatterings that are
forced may range from 0 up to any number.  Normally, only the first
scattering is forced, but if one is interested in the effects of
higher-order scattering, a larger number may be motivated.  If the
$j$'th scattering is not forced, an interaction optical depth is drawn
according to equation~\ref{equation-draw-optical-depth}.  The ray is
then propagated through the medium until it either leaves or reaches
the interaction optical depth.  If it reaches the interaction optical
depth, it is scattered.  The intensity of the ray after the interaction
is then 
\begin{equation}
I_{ i , j } = aI_{ i , j - 1 } \> ,
\end{equation}
 while the absorbed luminosity is 
\begin{equation}
A_{ i , j } = ( I_{ i , j } - I_{ i , j - 1 } ) n = ( 1 - a ) n I_{ i ,
j - 1 } \>.
\end{equation}
 The flux resulting from the intensity scattered in the direction of
the observer is 
\begin{equation}
F_{ i , j } = n I_{ i , j } e^{ - \tau_{ i , j }^{ \rm { obs } } }
\Phi_s ( \unitvec { k_{ i , j - 1 } } , \unitvec { k_{ i , j }^{ \rm {
obs } } } ) \frac { 1 } { d_{ i , j }^2 } \>.
\end{equation}

This process is repeated until the ray either leaves the volume or
until the intensity of the ray drops below some threshold, $I_{ \rm {
min } }$, set by the user.  To avoid expending computational resources
tracking rays with very low intensity that will not contribute
significantly to the results, the ray is dropped once its intensity
drops below $I_{ \rm { min } }$.  However, to avoid violating energy
conservation, this cannot be done in all cases. Instead, the ray is
given some probability $P_{ RR }$ of survival, and a random number is
drawn to see if the ray survives.  If it does, its intensity is
increased by a factor $1 / P_{ RR }$ and the ray continues to be
tracked.  If not, the ray is terminated.  This scheme, known as
``Russian roulette'', ensures that energy conservation, on average, is
preserved.

\subsection{Multiple Scattering Components}

It is possible to define an arbitrary number of scattering components
in every grid cell, for example if there are two different types of
dust with radically different scattering properties, distributed
differently.  In this case, the optical depths in the formulae above
refer to the sum of the optical depths of all the components.  When an
interaction takes place, the component responsible for the scattering
event is drawn randomly with a probability proportional to the
opacities of the components in the grid cell where the interaction
occurs.  The scattering is then performed using the albedo and
scattering phase function of this component.

It should again be pointed out that this procedure applies to the
transfer of radiation through the medium, for which only aggregate
quantities is necessary, and not to the determination of grain
temperatures.  For example, if the medium contains both dust with
Milky-Way-like and Small-Magellanic-Cloud-like properties, both of
which represent a \emph{distribution} of grain sizes and compositions,
this procedure is used when calculating the transfer of radiation
through this medium.  Only if the grain temperature distributions,
which are dependent on the full set of grain sizes and compositions, is
to be calculated is it necessary to consider each grain size
separately.

\subsection{Output Images}

The rays that are calculated to make it to the observer are projected
through a virtual ``pinhole camera'' onto an image plane.  These
cameras can be placed at arbitrary points.  (However, when cameras are
placed at a position where emission or scattering can occur, the noise
in the images will increase drastically due to the infinitely large
contributions from events occurring infinitesimally close to the camera
position.  This problem is known as the ``infinite variance
catastrophe'' \citep{mcprimer}, but is not likely to occur in
astronomical simulations.)  Each camera has a specified field of view
and image array size.

The surface brightness of pixel $p$ in the image is then calculated as 
\begin{equation}
\Sigma_p = \frac { \sum_{ i , j } F_{ i , j } } { \Omega_{ \rm { pix }
} }
\end{equation}
 where the sum over $i$ (ray number) and $j$ (interaction number) only
includes those $F_{ i , j }$ whose point of origin project onto pixel
$p$, and $\Omega_{ \rm { pix } }$ is the surface angle subtended by the
pixel for the observer.  If the projected ray is outside the field of
view of the camera, the light is lost.

\subsection{Absorbed Luminosity}

The absorbed luminosity in grid cell $c$ is calculated as 
\begin{equation}
A_c = \sum_{ i , j } A_{ i , j } ,
\end{equation}
 where the sum over $i$ and $j$ only includes those ray interactions
that occur in cell $c$.  The total absorbed luminosity in a grid cell
equals the total luminosity reradiated by the dust, by energy
conservation.  However, if the dust grain temperature distribution and
the SED of the dust emission is to be calculated self-consistently, the
(wavelength-dependent) radiation field in the cell must be determined
\citep{rajadraine89}.  If the absorbed luminosity is known, the
radiation field $J_c$  in the cell can be calculated as 
\begin{equation}
\label{equation_absorbed_intensity} J_{ c } ( \lambda ) = \frac { A_c }
{ 4 \pi \rho_c \kappa_{ \rm { abs } } V_c } \> ,
\end{equation}
 where $\kappa_{ \rm { abs } }$ is the absorption opacity of the dust
and $V_c$ is the volume of the cell.  Because only absorption events
contribute to the signal, this method suffers from large Monte-Carlo
noise in regions where the number of interactions are few, for example
in highly refined cells with small volume.  In fact, because the
radiative-transfer algorithm used by \mcrx \ is so much more efficient
at getting signal to the cameras than the simplest Monte-Carlo
implementation, fewer rays need to be traced.  This means that, since
each ray interacts with the medium at most a few times, the number of
absorption events determining $A ( \lambda )$ in 
equation~\ref{equation_absorbed_intensity} is small  and the quantity
noisy. \mcrx \ uses another method, described by \citet{lucy99} and
\citet{niccolini03}, that takes advantage of the fact that, physically,
the radiation intensity is determined by the number of rays (photons)
traversing a volume, regardless of the probability of absorption.  In
this scheme, 
\begin{equation}
\label{equation_traversed_intensity} J_{ c } = \frac { \sum_{ i , j }
\Delta \ell_{ i , j , c } n I_{ i , j } } { 4 \pi V_c } \> ,
\end{equation}
 where $\Delta \ell_{ i , j , c }$ is the path length during which the
$i$'th ray, after the $j$'th interaction, passes through cell $c$. 
Since many more rays pass through a given cell than are absorbed in it,
this method has superior accuracy.  The only complication is in the
case of forced scattering.  In this case the ray intensity in the cells
traversed before the forced scattering takes place is $I_{ i , 0 }$,
but the part of the ray that leaves the medium without interaction also
has to be taken into account.  That part of the ray has lower
intensity; it gives a contribution to $J_{ c }$ corresponding to $I_{ i
, 0 } e^{ - \tau_{ i , 0 }^e }$ in the cells traversed after the
interaction point.

\subsection{Polychromatic Ray Tracing}
 \label{polychromatic_section}

Looking back at the preceding sections, it is possible to identify the
points where wavelength dependence poses conceptual problems to a
procedure where all wavelengths are included in every ray.  Clearly,
emission of rays and calculation of the flux reaching the observer
either directly (Equation~\ref{equation-direct-flux}) or from a point
of scattering (Equation~\ref{equation-scattered-flux}) pose no
problems: these formulae are analytic calculations that can be
performed for any number of wavelengths simply by replacing the
quantities $I_i$, $\tau_{ i , j }^{ \rm { obs } }$ and $a$ with vectors
of numbers.  Polychromatic calculations of the direct flux has already
been done in the SKIRT code \citep{baesetal05heidelberg}.  The problems
arise where interaction optical depths and scattering directions are
sampled from the appropriate probability distributions, because these
probability distributions depend on wavelength.  For example, rays of
shorter wavelength will tend to travel shorter distance before
interacting, since the dust opacity generally increases towards shorter
wavelengths. This means an interaction point can only be drawn in a
statistically correct way for one wavelength at a time, and the same
objection applies to the scattering angle drawn using
equation~\ref{hg-draw-theta}. However, the ability to use biased
distributions opens up the possibility to compensate for the fact that
the probability distributions will only be correct for one wavelength. 
(This is known as ``path stretching''.)  The proper way of doing this
will now be examined.

As was derived in Section~\ref{section_drawing}, the probability
distribution function of where a ray interacts with the medium is 
\begin{equation}
\mathrm {d} P ( \tau ) = e^{ - \tau } \mathrm {d} \tau \>.
\end{equation}
 Suppose an interaction optical depth $\tau_{ \rm { ref } }$ is drawn
for some reference wavelength $\lambda_{ \rm { ref } }$. The
probability of another wavelength $\lambda$ interacting at the same
point is then 
\begin{equation}
\mathrm {d} P \left [ \tau ( \lambda ) \right ] = e^{ - \tau ( \lambda
) } \mathrm {d} \tau ( \lambda ) = e^{ - \frac { \tau ( \lambda ) } {
\tau_{ \rm { ref } } } \tau_{ \rm { ref } } } \left [ \frac { \tau (
\lambda ) } { \tau_{ \rm { ref } } } \right ] \mathrm {d} \tau_{ \rm {
ref } } \>.
\end{equation}
 The necessary biasing factor $w_\lambda$ is the ratio of the
probabilities at wavelengths $\lambda$ and $\lambda_{ \rm { ref } }$: 
\begin{equation}
\label{equation-tau-bias} w_\lambda = { \frac { P \left [ \tau (
\lambda ) \right ] } { P \left [ \tau_{ \rm { ref } } \right ] } } =
e^{ \tau_{ \rm { ref } } - \tau ( \lambda ) } \left [ \frac { \tau (
\lambda ) } { \tau_{ \rm { ref } } } \right ] \>.
\end{equation}
 To compensate for the biased probability distribution, the intensity
of the ray at different wavelengths at the point of interaction should
be multiplied by the weighting factor $w_\lambda$ before calculating
scattered or absorbed luminosity.

In cases where forced scattering is used, the probability distribution
from which interaction points are drawn is different, and so is also
the weighting factor.  The correct $w_\lambda$ when forced scattering
is used it is 
\begin{equation}
w_\lambda = e^{ \tau_{ \rm { ref } } - \tau ( \lambda ) } \left [ \frac
{ \tau ( \lambda ) } { \tau_{ \rm { ref } } } \right ] \left [ \frac {
1 - e^{ - \tau^e_{ \rm { ref } } } } { 1 - e^{ - \tau^e ( \lambda ) } }
\right ] \>.
\end{equation}

   Finally, the biased distribution of scattering angles must be
accounted for.  Compared to the optical depths, this is quite
straightforward: the probability of scattering into a certain direction
is given by the scattering phase function $\Phi_s ( \theta )$, so if a
scattering angle $\theta$ is drawn at the reference wavelength the
weighting factor which should be applied to the ray intensity after
scattering will be 
\begin{equation}
w_\lambda = \frac { \Phi_s ( \theta , \lambda ) } { \Phi_s ( \theta ,
\lambda_{ \rm { ref } } ) } \>.
\end{equation}

Energy conservation, in a statistical sense, must be maintained in the
polychromatic calculation; energy flux is the product of probability
and ray intensity, and any biasing scheme simply trades probabilities
for intensities.

The possibility of calculating all wavelengths simultaneously was noted
by \citet{juvela05}, who argued that it would not be advantageous since
the opacity is a strong function of wavelength and the large bias
factors necessary probably would result in increased errors.  It is
true that the errors \emph{for a fixed number of rays} probably would
increase for wavelengths where the dust opacity is very different from
what it is at the reference wavelength, but the differential errors
between different wavelengths are minimized.  This is clearly
illustrated in the example spectra calculated in
Section~\ref{section_polychromatic_tests}. 
Because every wavelength is uncorrelated in the monochromatic
calculation, the spectral shape becomes very noisy in regions of low
signal-to-noise, but this is not the case with the polychromatic
calculation.  (The use of ``correlated Monte Carlo'' for perturbation
analysis builds on the same principle, the stochastic effects are
minimized by correlating the random walks in the perturbed and
unperturbed cases.)  The increased noise at wavelengths far away from
the reference wavelength is alleviated by the fact that with the
polychromatic algorithm more rays can contribute to each wavelength. 
This is because the computational costs of tracing rays in \mcrx \ is
dominated by propagating rays from cell to cell during the random walk.
 As long as the vector operations for doing the calculation at many
wavelengths is not the dominant computational cost, the extra
wavelengths are obtained at little cost.

A great advantage of the polychromatic algorithm over the approach
using a set of discrete wavelengths is that the radiative-transfer
problem is truly solved for every wavelength; spectral features present
in the stellar emission are predicted properly, and the differential
attenuation between lines and continuum is accurately treated.  (This
can also be seen in the example spectra calculated in
Section~\ref{section_polychromatic_tests}.)  This is important for
predicting images and spectra of galaxies, the particular application
for which \mcrx \ was developed, where differential extinction between
different stellar populations can significantly affect spectral
features like the Balmer absorption lines.

Another advantage is that it facilitates the inclusion of kinematic
effects into the radiative transfer.  Kinematic effects are already
included in the SKIRT code \citep{baesetal03}, but then only a small
range in wavelengths, over which the dust characteristics is assumed to
be constant, is included in the calculation. Using the polychromatic
algorithm, the entire spectrum can be propagated as the accumulated
Doppler shift is tracked.  When the ray is finally projected on to the
camera, this Doppler shift can be included yielding a full spectrum
including Doppler broadening.

\subsection{Uncertainties \label{section_uncertainties}}

Because of the stochastic nature of the Monte-Carlo method, results are
subject to random sampling error.  If this error can be evaluated at
runtime, the number of rays used can be adapted to get the error below
some tolerance \citep{gordonetal01}, and even if this is not possible,
knowing the uncertainty in the results is important.

Estimating the uncertainty is conceptually straightforward.  The
quantity of interest (the flux in a pixel, for example) is the sum of
$N_V$ samples of a random variable $V$.  The variance in the estimate
of the sum of $N_V$ samples of a random variable is $N_V^2$ times the
variance of $V$, which in turn is estimated as 
\begin{equation}
\label{equation_standard_error} \sigma_V^2 = \frac { 1 } { N_V } \sum_i
v_i^2 - \left [ \frac { 1 } { N_V } \sum_i v_i \right ]^2 \>.
\end{equation}
 However, there is an important subtlety here.  A ``sampling'' of the
flux in a pixel consists of the shooting of one ray.  $v_i$ is the
total contribution made by the ray, so if one ray contributes flux in
the pixel both directly and through subsequent scatterings, the sum of
all these contributions is $v_i$.  This is important, because $v_i$ is
squared.  Squaring the contributions from direct, single-scattered,
etc., light separately will lead to a systematic underestimate of the
variance.  It should be pointed out that, for large $N_V$, the
estimated variance is insensitive to whether $N_V$ is taken to be the
total number of rays or just the number of rays with nonzero
contributions to the pixel.  (In general, most rays will not contribute
to the flux in a given pixel.)

\begin{figure}  \begin {center} \includegraphics*[width=0.95\columnwidth]{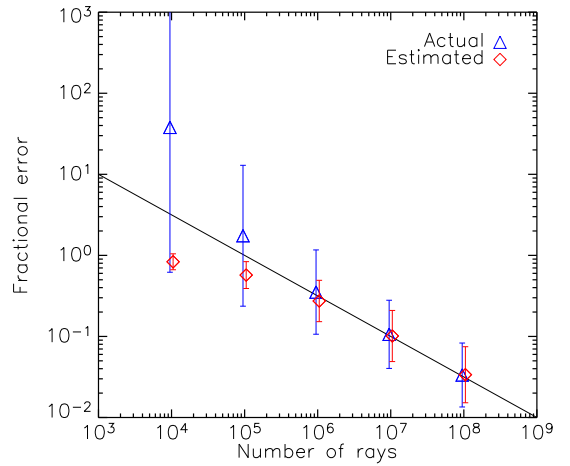}  \end {center} \caption{ \label{figure_variance}    The fractional errors (standard deviation/value) of the pixel values   in the image shown in Figure~\ref {wg96image} for different numbers   of rays, showing the convergence of the results.  The uncertainties   were estimated both using equation~\ref{equation_standard_error}   (red diamonds) and as the empirically determined variance from 30   runs (blue triangles).  The error bars show the $1\sigma$   logarithmic spread in the distribution of pixels, and the solid line   indicates the theoretically expected $\sqrt{N}$ convergence. For   small numbers of rays, equation~\ref{equation_standard_error}   severely underpredicts the actual variance.  (The standard deviation   estimated with equation~\ref{equation_standard_error} is bounded by   the quantity itself in cases where only one ray contributed to the   result.)  For increasing numbers of rays, the estimates converge   towards the empirically determined values, but does so   systematically from below. There is a tendency for the results   obtained with the ``next-event estimator'' to be dominated by rare,   large contributions, implying that unless the problem phase space is   well sampled, the estimated uncertainty will always be too small.  } \end{figure} 

This formula for estimating the uncertainty has been tested on the
pixel-by-pixel brightness in one of the cases of the clumpy scattering
medium of \citet{wittgordon96} against which the code is tested in
Section~\ref{section-WG96}. This test is shown in
Figure~\ref{figure_variance}. For different numbers of rays, the
estimated variances in the pixel values from
equation~\ref{equation_standard_error} were compared against the
empirically determined variances from 30 runs with different random
numbers.  For small numbers of rays, there are many pixels which have
one or no contribution, and the variance is severely underestimated. 
As the number of rays increases, the variances estimated using
equation~\ref{equation_standard_error} converge toward the empirically
determined values, but does so from below.  This emphasizes an
important fact about variance estimates using the next-event estimator:
the results tend to be dominated by rare, large contributions, so
unless the problem phase space has been thoroughly sampled, the true
variance will be underestimated.  (In fact, if scatterings can take
place close to the camera, the true variance is not even finite, as
previously mentioned.)

As seen above, it is necessary to save not only the quantity itself but
also the sum of the squared contributions to determine the uncertainty.
 This increases the amount of data to keep track of by a factor of two,
and makes the parallelization of the algorithm more complicated.
Because of this,  the uncertainty is normally not estimated in the
galaxy merger simulations (The outputs from one run are already
frequently larger than 1\,Gb).

\subsection{Clumpy Dust Distributions}

Many authors have studied the effects of clumpy distributions of dust
\citep{wittgordon96, wittgordon00, bianchietal00} and reached the
conclusion that it can profoundly affect the attenuation of radiation
for a given mass of dust.  The general tendency is for clumping to
decrease attenuation and reddening by allowing radiation to escape
through optically-thin lines of sight, unless the sources themselves
are embedded in the clumps, in which case the attenuation can be higher
than the corresponding homogenous scenario.

In \mcrx , dust is assumed to be uniformly distributed within each grid
cell.  There is no additional clumping assumed beyond what is actually
present in the adaptive grid.  In our hydrodynamic galaxy merger
simulations there are large-scale inhomogeneities, such as spiral
structure, present, but the resolution (and the physics incorporated,
for that matter) is too coarse to resolve e.g. individual molecular
clouds.  The adaptive grid structure could be used to put in artificial
clumping on yet smaller scales, but the computational requirements
would be prohibitive.  One solution would be to incorporate clumping
through a sub-grid analytical approximation, such as the ``mega-grains
approximation'' \citep{neufeld91, hobsonpadman93, varosidwek99}; large
clumps can be treated as enormous dust grains, with an effective albedo
and scattering function.  Clumps of dust can then be added within the
existing framework as just another source of scattering.  If the
sources of emission are located within the clumps, however, this
approach will not work.  This can instead be accommodated by changing
the characteristics of the emitted radiation before injecting the rays
into the grid.

\section{Auxiliary Calculations -- Radiative Transfer in Hydrodynamic
Simulations }
 \label{auxiliary_calculations_section}

In section~\ref{radiative_transfer_section}, the core
radiative-transfer algorithm of \mcrx \ was described.  However, in
order to generate images and spectral energy distributions from
hydrodynamic N-body simulations, the purpose for which the code was
written, additional steps are necessary.  These extra steps are
described in this section.

The general procedure is as follows: first, hydrodynamic simulations
are used to generate the geometry of the problem, e.g. merging
galaxies.  A number of snapshots at different time steps are saved, and
for each of these a series of preparatory steps is performed. First, a
stellar population synthesis model is used to calculate the SEDs of the
emitting sources.  Second, the adaptive grid needed for the
radiative-transfer calculations is generated.  Third, the
radiative-transfer calculations are done. (Work is currently being done
to integrate the polychromatic version of \mcrx \ into this framework. 
The hydrodynamic simulations processed so far have used to the
monochromatic version of \mcrx .) Finally, a post-processing step is
done, where the full SED is calculated by interpolating over
wavelength.

\subsection{The Hydrodynamic Simulations\label{section_simulations}}

The hydrodynamic simulations have been described in detail elsewhere
\citep{tjthesis, pjetal05attn, coxetal05methods}. In particular, the
specific galaxy models which are being used for the galaxy merger
simulations are described in \citet{pjetal05attn} and will not be
discussed here, but in order to provide a context and define the
quantities used in the radiative-transfer calculations, a brief
overview of the technique used will be given.

The simulations are done using GADGET \citep{springeletal01,
springel05g2}, a Lagrangian Smooth Particle Hydrodynamics (SPH) code.
The galaxies are initially modeled as a disk of stars and gas, a
stellar bulge, and a dark-matter halo.  The stars and dark-matter
particles are collisionless and only feel the force of gravity.  The
gas particles are also subject to hydrodynamic forces.  A collisionless
particle $i$ is characterized by its mass $m_i$ and its gravitational
softening length $r_i$.  A gas particle has, in addition, an associated
SPH smoothing length, $h_i$, which indicates the size of the region
over which the hydrodynamic quantities associated with the particle are
averaged.  The smoothing length is determined by the distance to
neighboring gas particles, i.e. by the gas density, such that the
resolution is higher in high-density regions and lower in low-density
regions where the particles represent a very smoothed-out gas density
field.

During the simulation, the star-formation rate of each gas particle is
calculated according to a ``Schmidt law'', 
\begin{equation}
\label{Schmidt_law} \frac { \mathrm {d} \rho_\star } { \mathrm {d} t }
\propto \rho_{ \rm { gas } }^{ 1.5 } \>.
\end{equation}
 As stars form, gas is transformed into collisionless matter.  In the
simulations, this is implemented in a stochastic sense
\citep{springelhernquist03} in which each gas particle spawns a number
of stellar particles with a probability consistent with the calculated
star-formation rate.  These ``new star'' particles represent
associations of single stellar populations, though their mass,
typically $10^6 \Msun$, is larger than most observed young star
clusters.  In addition to the quantities associated with all
collisionless particles mentioned earlier, new star particles are also
characterized by a formation time $t_{ { \rm f } , i }$ and a
metallicity $Z_i$, which is the metallicity of the gas particle from
which it is spawned.

Associated with star formation is supernova feedback, whereby energy
from supernovae is deposited into the interstellar medium.  This energy
heats and pressurizes the gas and stabilizes it from further
gravitational collapse.  Including feedback is crucial for the
stability of the simulations, but it is a complicated subject and many
different approaches to implementing it exist.  Here, it will only be
mentioned that our simulations contain feedback and that a significant
effort has gone into constraining this part of the simulations
\citep{coxetal05methods}. Feedback from AGN could also affect the gas. 
This has been included in GADGET simulations \citep{springeletal04},
and these simulations can also be processed by \mcrx .

A third consequence of star formation is chemical enrichment.  Since
the goal is to simulate the effects of dust, tracking metal production
is naturally a topic of interest as the amount of metals available will
affect the amount of dust present.  In the simulations, metal
production is implemented using an ``instantaneous-recycling'' scheme:
stars formed are assumed to instantly become supernovae, and the metals
produced are put back into the gas phase of the particle. Every gas
particle is, in addition to the quantities mentioned earlier,
characterized by a metallicity $Z_i$.  This scheme, while simple, has
several drawbacks: First, metals do not diffuse from the gas particle
from which they were made, and if this particle is completely turned
into stars, all the metals are incorporated into the stars and lost
from the ISM.  Second, while metals are recycled in supernovae, gas is
not.  In reality, a stellar population returns a non-negligible
fraction of its mass to the gas phase, due to supernovae and stellar
winds.  Third, while instantaneous recycling may be a reasonable
approximation for Type II supernovae, it is surely not for Type Ia
supernovae which are believed to explode at least several hundred
million years after the stars are formed.  Improved models for metal
enrichment in GADGET have been developed \citep{scannapiecoetal05}, but
our simulations have so far not used these.

\subsection{Calculation of Stellar SEDs}

After the hydro simulations have been completed, the first step is to
calculate an SED for the stellar particles in each simulation snapshot.
 In this work, the SEDs used are from Starburst99
\citep{leithereretal99} but are subsampled to 510 wavelength points in
order to minimize file size.  (Note that this is not the much smaller
number of wavelengths for which the monochromatic radiative-transfer
calculations are done.)  The calculation of the stellar SED is trivial
since the assumption is that stellar particles represent single stellar
populations, so one simply has to pick an SED with the age and
metallicity of the particle.

For stars present at the start of the simulation, i.e. the disk and
bulge components of the merging galaxies, the star-formation history
and metallicity distribution must be specified as input parameters.
Typically, the bulge stars are assumed to have formed in an
instantaneous burst a relatively long time ago, while the disk has had
an exponentially declining star-formation rate starting at the time of
bulge formation and leading up to the start of the simulation, but any
choice can be made.  Formation times consistent with these assumptions
are then drawn randomly for the individual particles.

The physical size of the region over which the particle luminosity is
distributed is set to some fixed value.  Typically, the gravitational
softening length $r_i$ is used.

\subsection{Creating the Adaptive Grid \label{adaptive_grid_section}}

The amount of dust is based on the amount of metals present in the
galaxy simulations. As was described in
section~\ref{radiative_transfer_section}, the ray tracing is done on a
grid, so it is necessary to transform the density field described by
the particles onto a grid.  In order to be able to resolve the small
high-density regions in the simulations, while still covering the large
volume over which the interaction takes place, this grid is adaptive
\citep{wolfetal99,kurosawahillier01, harriesetal04,
stamatelloswhitworth05}.  The grid structure is based on a regular
Cartesian grid, in which grid cells can be recursively refined by
subdividing them into $2^3$ subcells,  leading to an octtree-like
memory structure.

Grid construction proceeds as follows.  First, the base grid is
created.  This grid covers the entire extent of the geometry to be
simulated, and typically has $10^3$ grid cells.  Each of these cells is
then recursively subdivided until $l_c < \min_i ( h_i ) c$, i.e. the
cell is smaller than the sizes of all particles contained within it.
This strategy uses the information contained in the SPH smoothing
length to determine the smallest scale which possibly contains
structure.  The constant $c$ is a fudge factor, typically chosen to be
2, since this is roughly the region over which the particle contributes
density. (The SPH smoothing kernel extends to $2 h_i$.)

Once the refinement step is completed, the mass of metals contained in
the particles is projected onto the grid cells using the SPH smoothing
kernel as the radial density profile.  For the projection, the
piecewise polynomial kernel in \citet{hernquistkatz89} is used.
Performing this three-dimensional integral is time-consuming, and
therefore a tabulated version is used.  In the cases where the particle
is much larger than the grid cell, the density associated with the
particle is assumed to be constant over the extent of the grid cell,
eliminating the need for the integration.

At this point, the grid describes the spatial distribution in the
simulations but not necessarily in the most efficient way.  Unlike in,
for example, an adaptive-mesh hydrodynamics code, the necessary size of
the grid cells does not depend on the \emph{magnitude} of the density,
only on its inhomogeneity.  There is no size scale that has to be
resolved, as long as the problem is described accurately.  (This is not
true for the determination of the radiation intensity, since this can
obviously change even if the dust density is perfectly uniform. 
Unfortunately, there is no local criterion for determining the
resolution required to resolve the radiation field without actually
solving the radiative-transfer problem.) Given this fact, it is
desirable that cells whose quantities are sufficiently uniform be
unified into one larger cell, as the ray tracing takes approximately
constant time per grid cell traversed.  Subcells are unified and
replaced with their ``parent'' cell as long as the following criterion
is fulfilled: 
\begin{eqnarray}
\label{equation_gas_tolerance} \left ( \frac { \sigma ( \rho_{ \rm {
met } } ) } { \left < \rho_{ \rm { met } } \right > } < { \mbox { tol }
\! }_{ \rm { met } } \right ) & { \mbox { OR } } & \left ( \sigma (
\rho_{ \rm { met } } ) < \frac { V_{ \rm { grid } }^{ - 2 / 3 } } {
\kappa N } \right ) \> ,
\end{eqnarray}
 where the average ($\left <...\right >$) and standard deviations
($\sigma$) are calculated over the subcells, $V_{ \rm { grid } }$ is
the volume of the entire grid, $N$ is the number of Monte-Carlo rays to
be traced, and $\kappa$ is an opacity ($\mathrm {d} \tau / \mathrm {d}
x$).  Two different kinds of criteria are used: The first is a relative
one; the standard deviation of the quantity over the subcells divided
by the average quantity (the value the unified cell would have) must be
less than a specified tolerance ($\rm { tol }_{ \rm { met } }$) for
unification to be allowed.  This ensures that inhomogeneities are
resolved.  The second criterion is an absolute one; the standard
deviation of the subcells must be smaller than some value.  The idea is
that if the difference in the quantity resulting from unification is so
small that not even one Monte-Carlo ray will be affected by such a
change, we can unify the cells regardless of how large the fractional
deviation is. This ensures resources are not wasted on resolving
regions that are so sparse they will not matter to the results anyway.

If we have $N$ rays traversing the volume $V_{ \rm { grid } }$, the
number traversing a subvolume $v$ will be $n = N ( v / V_{ \rm { grid }
} )^{ 2 / 3 }$, if we assume a uniform and isotropic distribution of
rays.  The number of rays that will interact in the subvolume is 
\begin{equation}
n_s \approx n \Delta \tau = n \kappa \rho \Delta \ell = \frac { N
\kappa m } { V_{ \rm { grid } }^{ 2 / 3 } } \> ,
\end{equation}
 interestingly independent of $v$.  From this one can conclude that a
deviation in mass can be accepted without affecting a single scattering
if 
\begin{equation}
\Delta m < \frac { V_{ \rm { grid } }^{ 2 / 3 } } { \kappa N } \>.
\end{equation}
 Since this result is valid for a uniform and isotropic distribution of
rays, which is a questionable assumption, it is obviously not to be
trusted in detail.  It does however provide a scaling, and to ensure
cells are not excessively unified, $N$ is usually taken to be at least
$10 \times$ the number of Monte-Carlo rays actually used.

The grids used for the galaxy merger simulations are constructed with
$\mbox { tol } \!_{ \rm { met } } = 0.1$, $N = 10^7$ and $\kappa = 3
\times 10^{ - 5 } \rm { \kpc }^2 \! / \Msun$.  With these parameters,
the grids typically have 30k to 100k cells and cover a volume of $( 200
\rm { \kpc } )^3$, with a maximum of 10 subdivisions. This is the
equivalent resolution of a ${ 10240 }^3$ uniform grid.  Given the
run-time scaling in Section~\ref{section-run-time}, the use of an
adaptive grid allows the simulations to complete in 2\% of the time
that would be required for a uniform grid. Clearly this dynamic range,
which matches the dynamic range in the hydro simulations themselves,
would not be possible without an adaptive grid.

It is also worth mentioning that although this description was based on
building the grid from the information in the hydro simulations, the
code includes a generic ``grid factory'' class which can be used to
build the grid using density fields from any source.  In addition, the
grid can contain diffuse emission and the refinement also take the
inhomogeneities in emissivity into account.

\subsection{The Radiative-Transfer Stage}

The radiative-transfer algorithm as described in detail in
section~\ref{radiative_transfer_section} is repeated for a number of
wavelengths in order to get an idea of the appearance of the system at
all observed wavelengths, and to determine the total amount of energy
absorbed by the dust.  The assumption is then that since the dust
properties change smoothly with wavelength, the results can be
interpolated over wavelength to give a full SED.

The radiative-transfer stage is done in two phases.  First, a run
without considering any dust effects is done.  The entire SED (all
wavelengths at once) is simply propagated from randomly drawn emission
sites directly to the cameras using
Equation~\ref{equation-direct-flux}, giving images of the object.  This
is possible since without dust there is no wavelength-dependent part to
this process.  These ``dust-free'' images serve as a basis for the
later interpolation over wavelength.  Second, the full radiative
transfer including scattering and absorption is done for a number of
wavelengths from the far-UV to the near-IR.  This range includes
practically all the energy emitted by stars, so energy outside of this
range is ignored.  The number of wavelengths used is a trade-off
between accuracy and computation time.  For the galaxy simulations, 19
wavelengths, shown in Table~\ref{wavelength_table}, have been used. 
They were determined through a fitting procedure where the error in the
fraction of the luminosity absorbed by dust was minimized against a
test run using 100 wavelengths, and are particularly dense around the
$2200 \Angstrom$ feature in the Milky-Way dust extinction curve to
ensure that this region of the spectrum is calculated accurately.

\begin{table}   \begin {center} \begin{tabular}{l@{\hspace{10mm}}l} Wavelength/m & Wavelength/m (cont.) \\ \hline $2.09\cdot 10^{-8}$ & $3.45\cdot 10^{-7}$ \\

$5.03\cdot 10^{-8}$ & $3.94\cdot 10^{-7}$ \\

$7.18\cdot 10^{-8}$ & $4.15\cdot 10^{-7}$ \\

$9.06\cdot 10^{-8}$ & $4.86\cdot 10^{-7}$ ($\hbeta$)\\

$1.38\cdot 10^{-7}$ & $4.88\cdot 10^{-7}$ \\

$1.56\cdot 10^{-7}$ & $6.56\cdot 10^{-7}$ ($\halpha$)\\

$1.83\cdot 10^{-7}$ & $6.58\cdot 10^{-7}$ \\

$2.04\cdot 10^{-7}$ & $9.00\cdot 10^{-7}$ \\

$2.39\cdot 10^{-7}$ & $1.80\cdot 10^{-6}$ \\

$2.54\cdot 10^{-7}$ & $4.99\cdot 10^{-6}$ \\

$2.67\cdot 10^{-7}$ \\ \hline \end{tabular} \end {center} \caption[Wavelengths for the radiative-transfer calculation]{ \label{wavelength_table} The wavelengths for which the radiative-transfer calculation is done. } \end{table}

The assumption of smoothness as a function of wavelength is not valid
if one considers line emission such as $\halpha$.  While the dust
properties do not change over the line wavelength, the emission
properties change abruptly; emission lines come from star-forming
regions, which are generally more deeply embedded within the dust than
are the sources contributing to the continuum around the line.  This
means that emission lines can have an attenuation many times larger
than the continuum around them, and as a consequence their dust
dependence must be calculated by specifically doing the radiative
transfer in the line.  Currently, the hydrogen lines $\halpha$ and
$\hbeta$ are included, as these are frequently used as star-formation
and dust indicators.

\subsection{Post-processing}

After the radiative-transfer calculation has been done, the results are
interpolated to give an SED with the wavelength resolution of the
stellar models.  For every pixel in the output images, the attenuation,
defined as the surface brightness in the image calculated including the
effects of dust divided by the surface brightness in the image not
including dust, is calculated for each of the radiative transfer
wavelengths.  This attenuation is then interpolated over all the
wavelengths in the stellar-model SED, and this interpolated function is
multiplied by the dust-free SED.  This procedure yields an SED which
has all the spectral features of the stellar model, with a large-scale
behavior determined by the dust attenuation.  The only exception to
this is when the attenuation is $\gg 1$, i.e. there is more light in
the images with dust than in those without.  This frequently happens
when light is scattered by dust in regions that contain no emission. 
In these cases, the attenuation is of questionable meaning, and the SED
is interpolated directly from the radiative-transfer results.  (Even in
the cases where the attenuation is $< 1$, the interpolated SED is an
approximation since it does not account for the small-wavelength
spectral features of light scattered into the line of sight, only for
those present in the direct light.  When \mcrx \ is updated to use the
polychromatic algorithm for the galaxy simulations, this will no longer
be an issue.)  These data cubes are then integrated over a number of
filter passbands to generate broadband images of the simulated object.

Also during the postprocessing stage, the total luminosity absorbed in
every grid cell is calculated by interpolating the absorbed luminosity
at the radiative-transfer wavelengths and integrating over all
wavelengths.  By energy conservation this is equal to the total
luminosity reradiated by the dust in the mid- and far-infrared, and
images of the reradiated luminosity are then generated by assuming zero
opacity and using Equation~\ref{equation-direct-flux}.

While there is no self-consistent calculation of the dust temperature
distribution and hence the SED of the radiation emitted by the dust, as
done by e.g. \citet{misseltetal01}, a rough idea of what the infrared
SED should look like, in the context of galaxies, is provided by the
templates of \citet{devriendtetal99}.  These templates provide the dust
emission SED of the galaxy as a function of its total infrared
luminosity, which makes it possible to estimate the total luminosity in
any given far-infrared passband.  It does not provide an estimate of
the spatial variations of the dust-emission SED, so constructing images
of the far-infrared emission can only be done assuming all cells have
the same dust-emission SED.  A self-consistent calculation of dust
temperature is a planned upgrade to the code.

Images are also generated for other quantities from the hydrodynamic
simulations.  The quantities imaged are mass density of stars, gas and
metals, star-formation rate, bolometric luminosity, mass- and
luminosity-weighted stellar age, and gas temperature.  These images
make it possible to correlate the emerging radiation and the dust
effects with physical quantities of the system on a pixel-by-pixel
basis.

\section{Tests}
 \label{test_section}

With any scientific code, it is crucial to test it against independent
results, either calculated analytically or using another (hopefully
correct) code.  The full radiative-transfer problem, with nonisotropic
scattering, absorption and arbitrary geometry is far too complicated
for analytic solutions, so the code has been tested against published
results obtained with other Monte-Carlo codes. 
Sections~\ref{section-W77} -- \ref{section-WG96} show comparisons with
simpler scenarios in which the geometry is well-defined and does not
factor into the uncertainty. Section~\ref{section_polychromatic_tests}
show how the polychromatic algorithm performs.  Finally, in
Section~\ref{section-P04}, \mcrx \ is used to calculate the 2D
benchmark problem from \citet{pascuccietal04}, which is also used to
demonstrate the relative efficiencies of the mono- and polychromatic
methods.

\subsection{Comparison to Witt (1977)}
 \label{section-W77}

\begin{figure*} \begin {center} \includegraphics[width=0.49\textwidth]{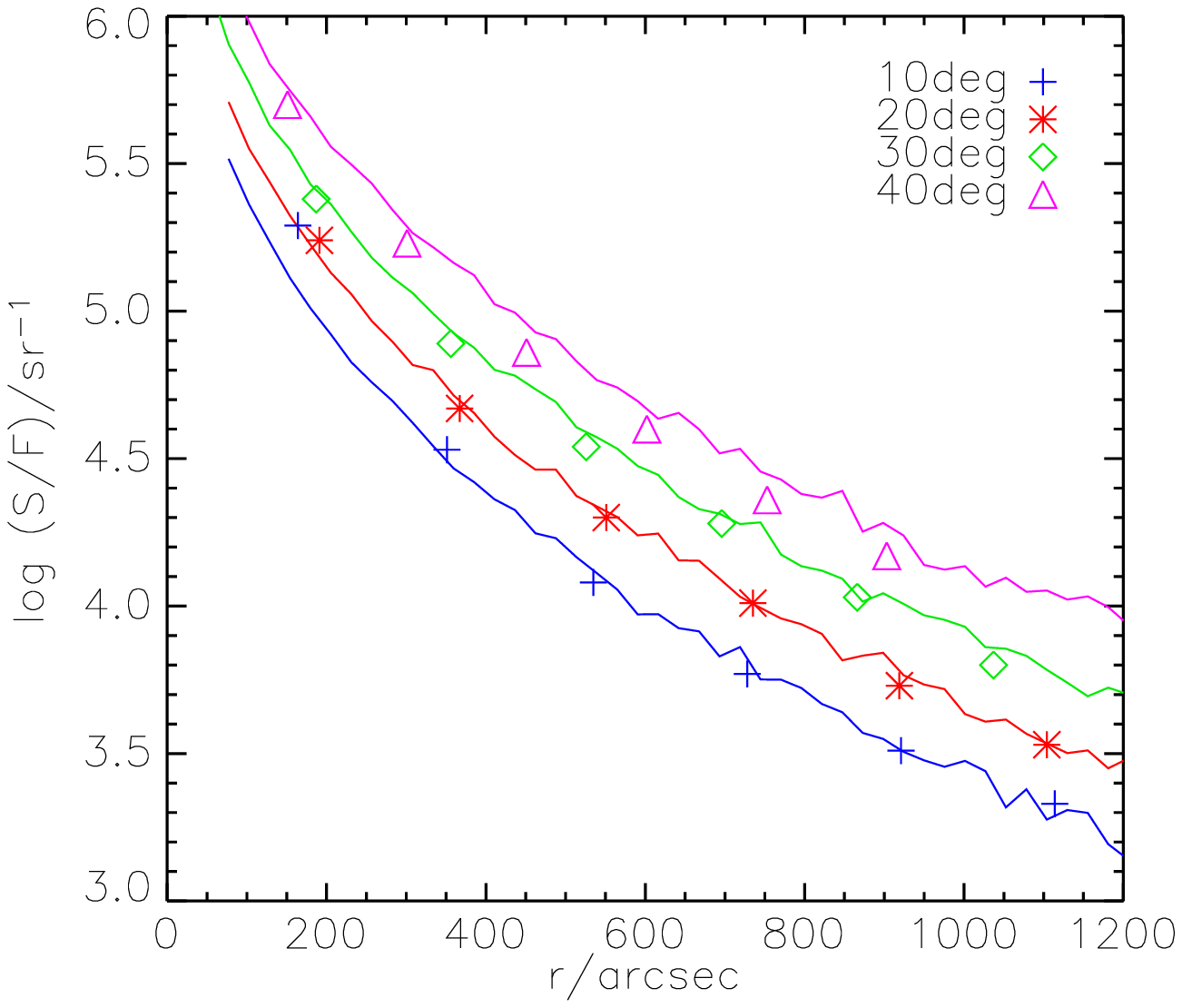} \includegraphics[width=0.49\textwidth]{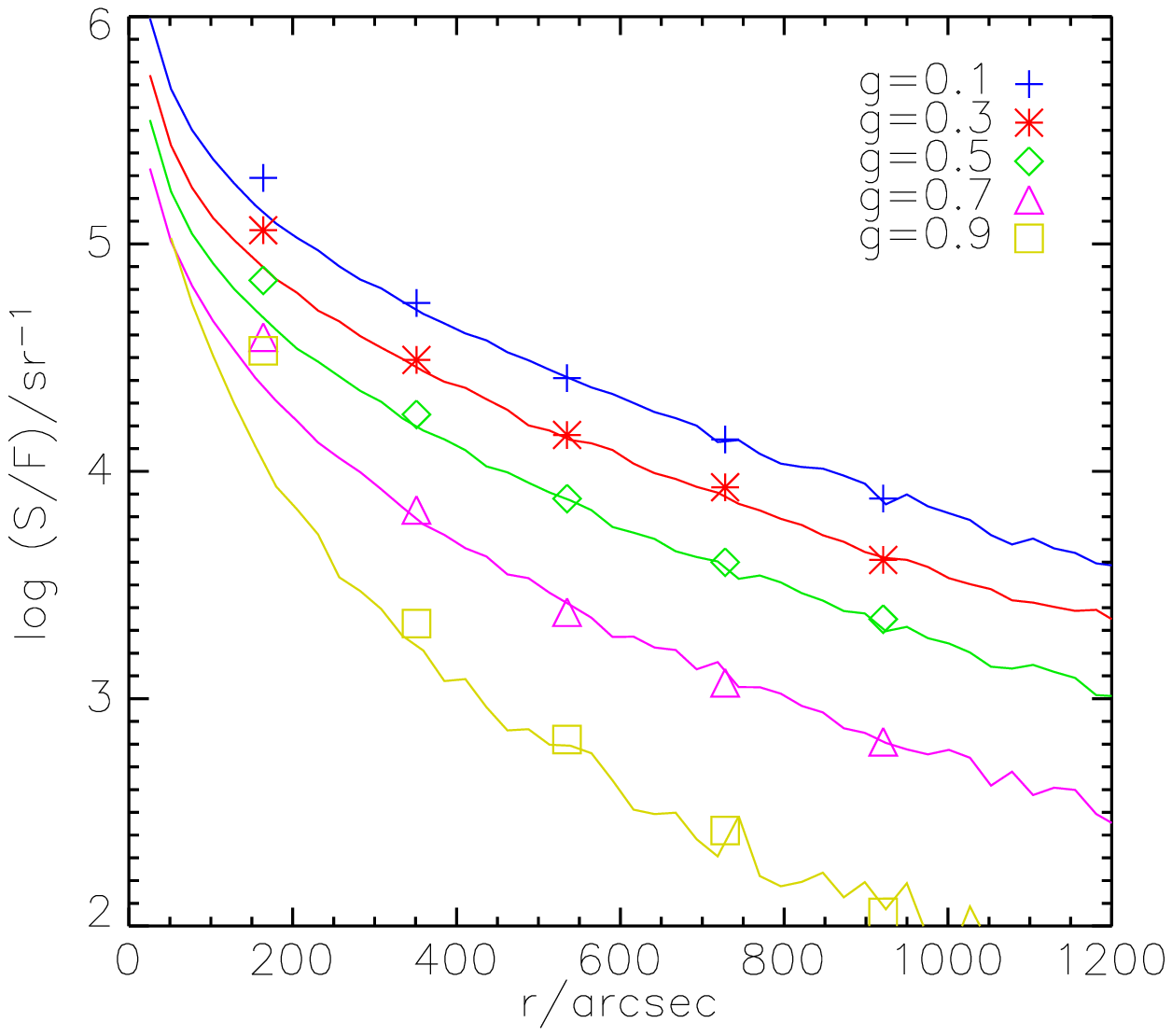}

\includegraphics[width=0.49\textwidth]{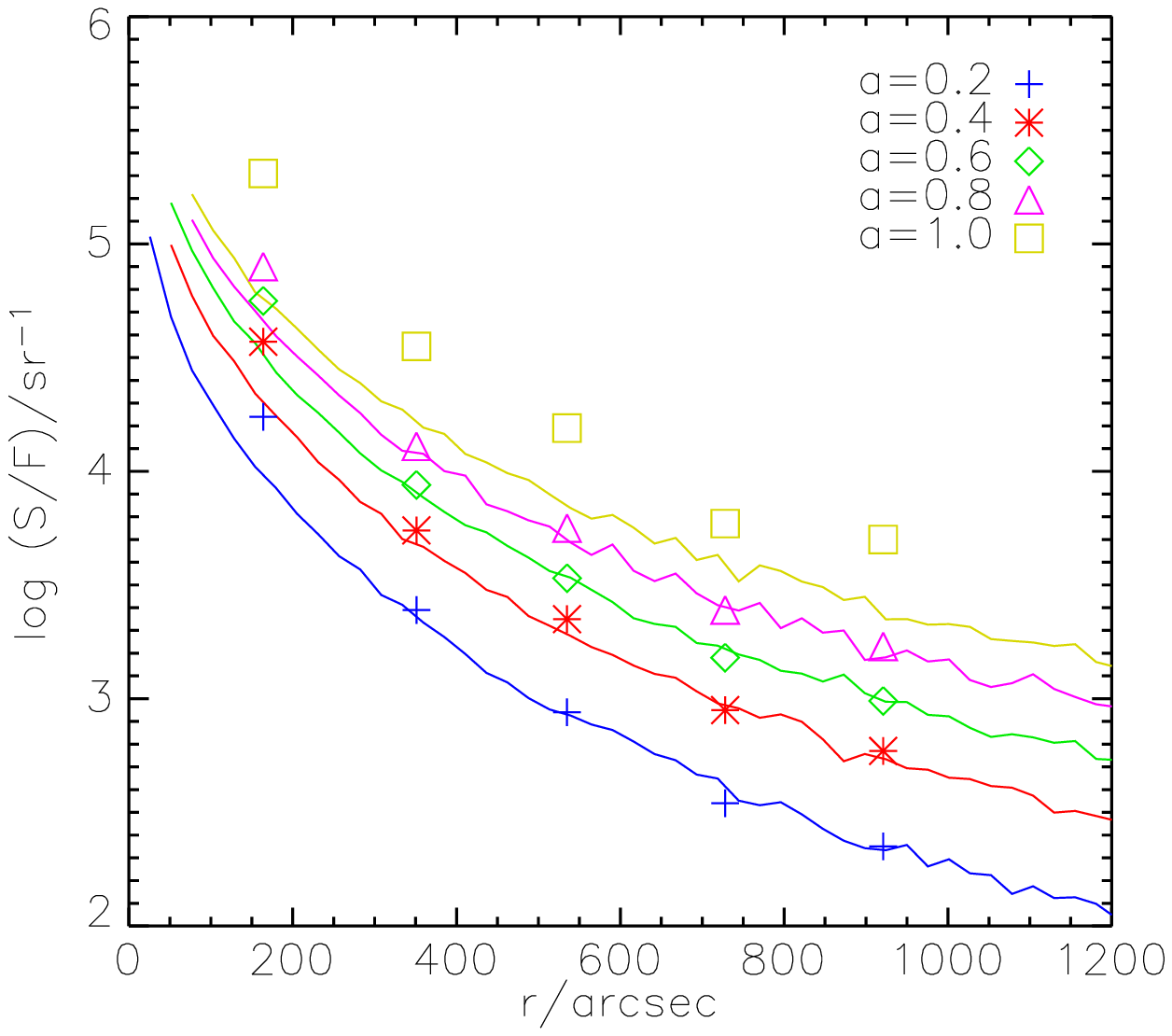} \includegraphics[width=0.49\textwidth]{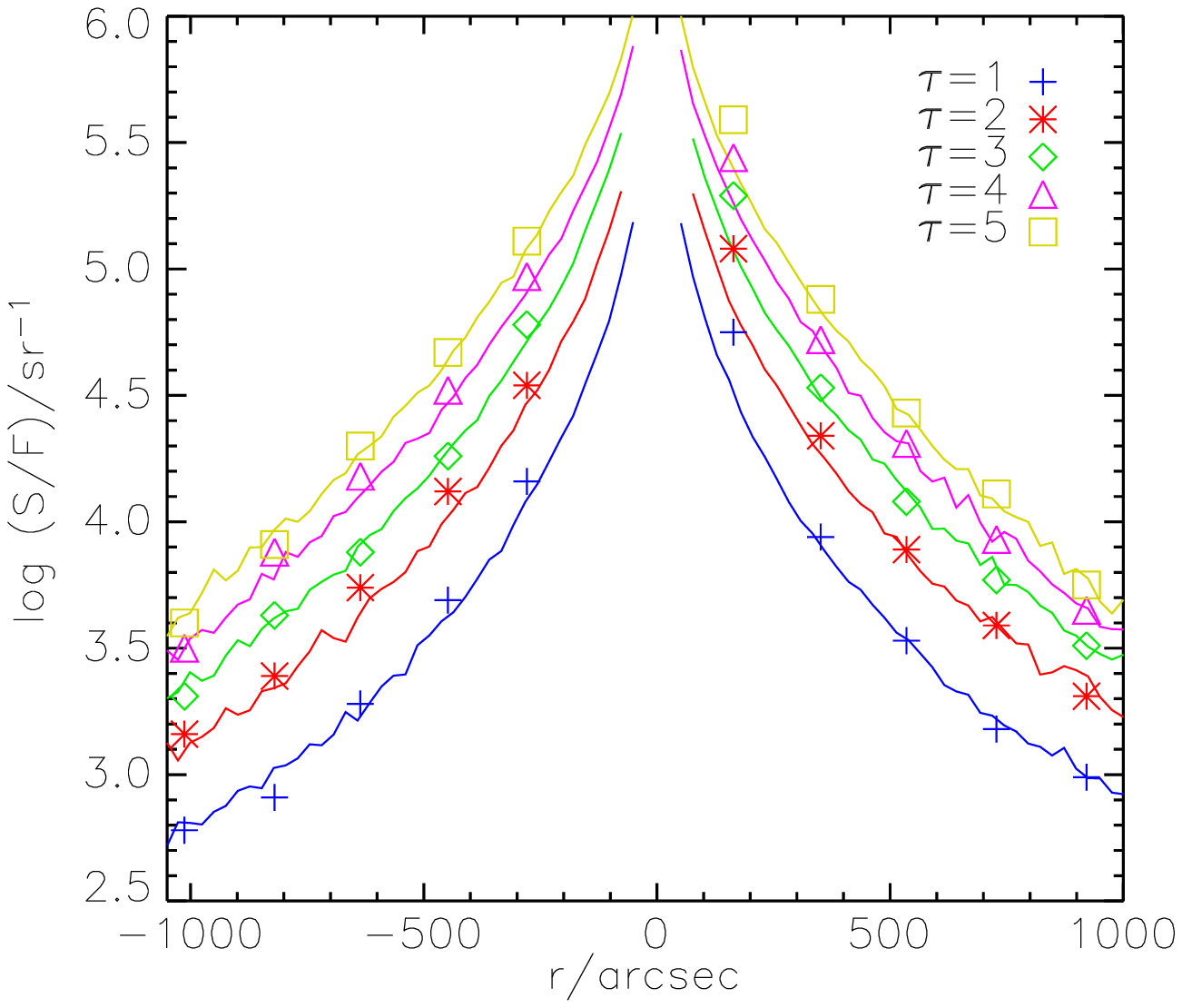} \end {center}  \caption{ \label{figure_w77} The radial brightness profiles of a nebula with an embedded star, in comparison with \citet{witt77III}. The plots show the surface brightness of the nebula, normalized by the flux of the star, as a function of the angular distance away from the star. The \mcrx\ results are shown as lines, while the W77c results (after multiplying them by 2, see text) are shown as points.   The four panels explore the dependence of the results on four different parameters: In the upper left, the dependence on viewing angle (W77c, Table~7); in the upper right, dependence on scattering asymmetry (W77c, Table~10); in the lower left, dependence on albedo (W77c, Table~11); and in the lower right, dependence on optical depth (W77c, Table~4). In general, the agreement is good except that \mcrx\ seems to systematically predict a less steep rise in brightness close to the star.  As explained in the text, this is consistent with the wide bins used by W77 along with the steep rise in surface brightness for small radii. There also appears to be a trend, in the upper left plot, for the \mcrx\ results to rise faster than the W77c data towards more oblique angles. The most seriously discrepant case is the $a=1$ case in the lower left, where W77c predicts a much higher surface brightness than that predicted by \mcrx. The origin of this discrepancy is unknown.  } \end{figure*}

In a series of papers,
\citet [] [hereafter collectively referred to as W77]  {witt77, witt77II, witt77III}
used a Monte Carlo radiative transfer code to calculate the emerging
surface-brightness distribution from a reflection nebula.  This work is
a suitable test for radiative-transfer calculations, because the
problem geometry is simple and the papers contain an extensive study of
how the surface-brightness distribution is affected by changing the
input parameters.  The nebula consists of a cylindrical, homogenous
slab of dust.  The star is located along the centerline of the
cylinder, either in front of the dust, or embedded within it.  The
surface-brightness distribution was then described by the quantity $S /
F$, where $S$ is the surface brightness of the nebula and $F$ is the
flux from the star, observed at the distance of $126 \pc$, as a
function of the angular distance from the star.  In \mcrx , the
adaptive grid was used to approximate the analytically defined cylinder
used by W77.  In practice, the results were insensitive to the degree
of grid refinement used.

A selection of the cases calculated by W77, varying the viewing angle,
grain albedo, scattering asymmetry, and optical depth of the nebula,
are presented in Figure~\ref{figure_w77}.  In general, the results
agree well, after a constant factor of 2 discrepancy in surface
brightness has been accounted for.  This discrepancy, in the sense that
the W77 results are too low, appears attributable to a normalization
error in the W77 results (A. Witt 2000, private communication), as it
has been confirmed by other codes (A. Watson 2000, private
communication).  There is a systematic trend where \mcrx \  predicts a
less steep rise in surface brightness close to the star, but this is
consistent with the expected effect from the wide radial bins used by
W77 in combination with the steep non-linear rise in surface brightness
towards the center.

The largest discrepancy is the $a = 1.0$ case in the lower left of
Figure~\ref{figure_w77}, where W77 predicts a surface brightness
significantly higher than \mcrx .  The origin of this discrepancy is
unknown, but unlike the W77 data, the \mcrx \ results appear to be the
extrapolation of the lower albedo data.

\subsection{Comparison to Watson \& Henney (2001)}

\begin{figure*} \begin {center} \includegraphics[width=0.49\textwidth]{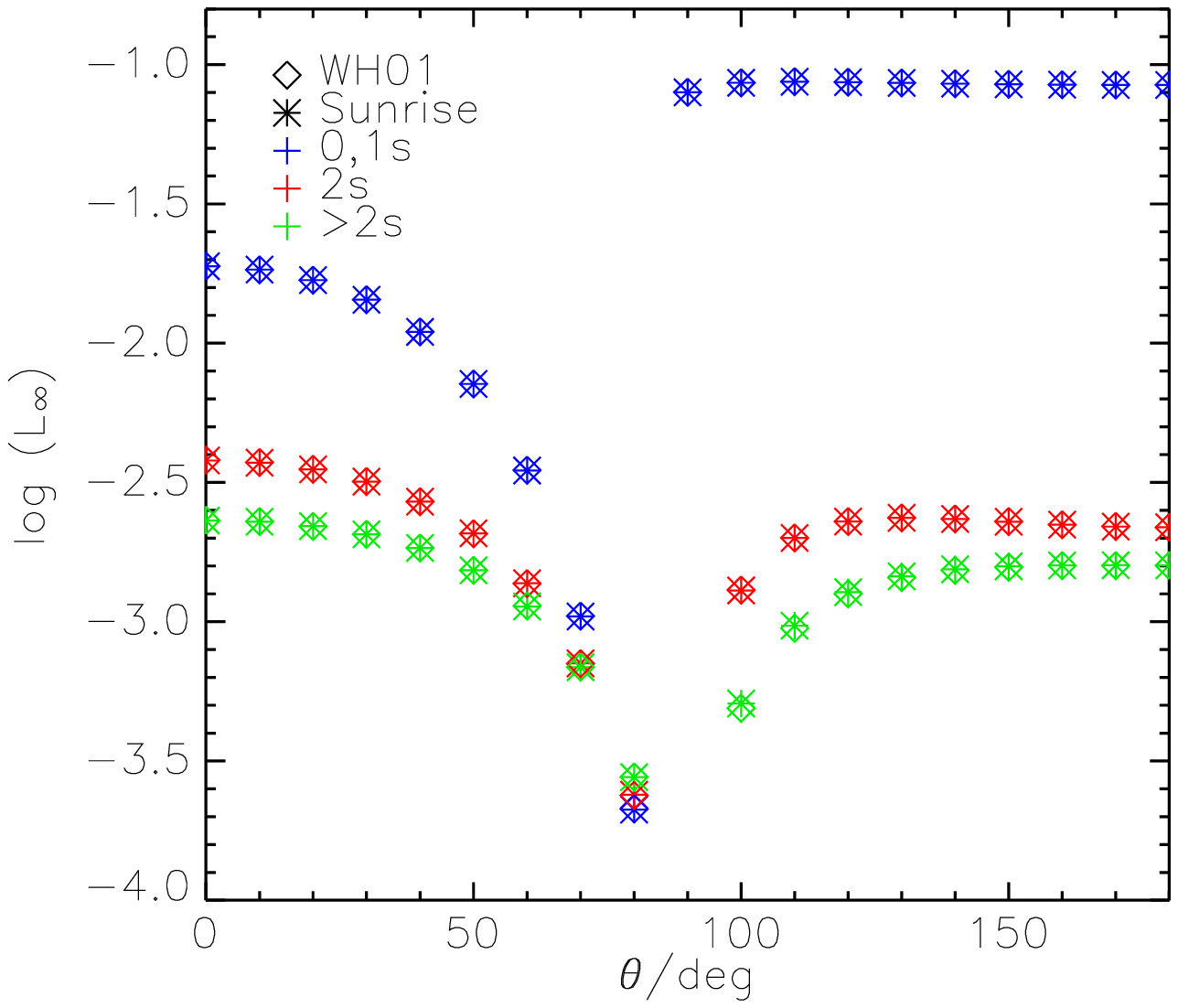} \includegraphics[width=0.49\textwidth]{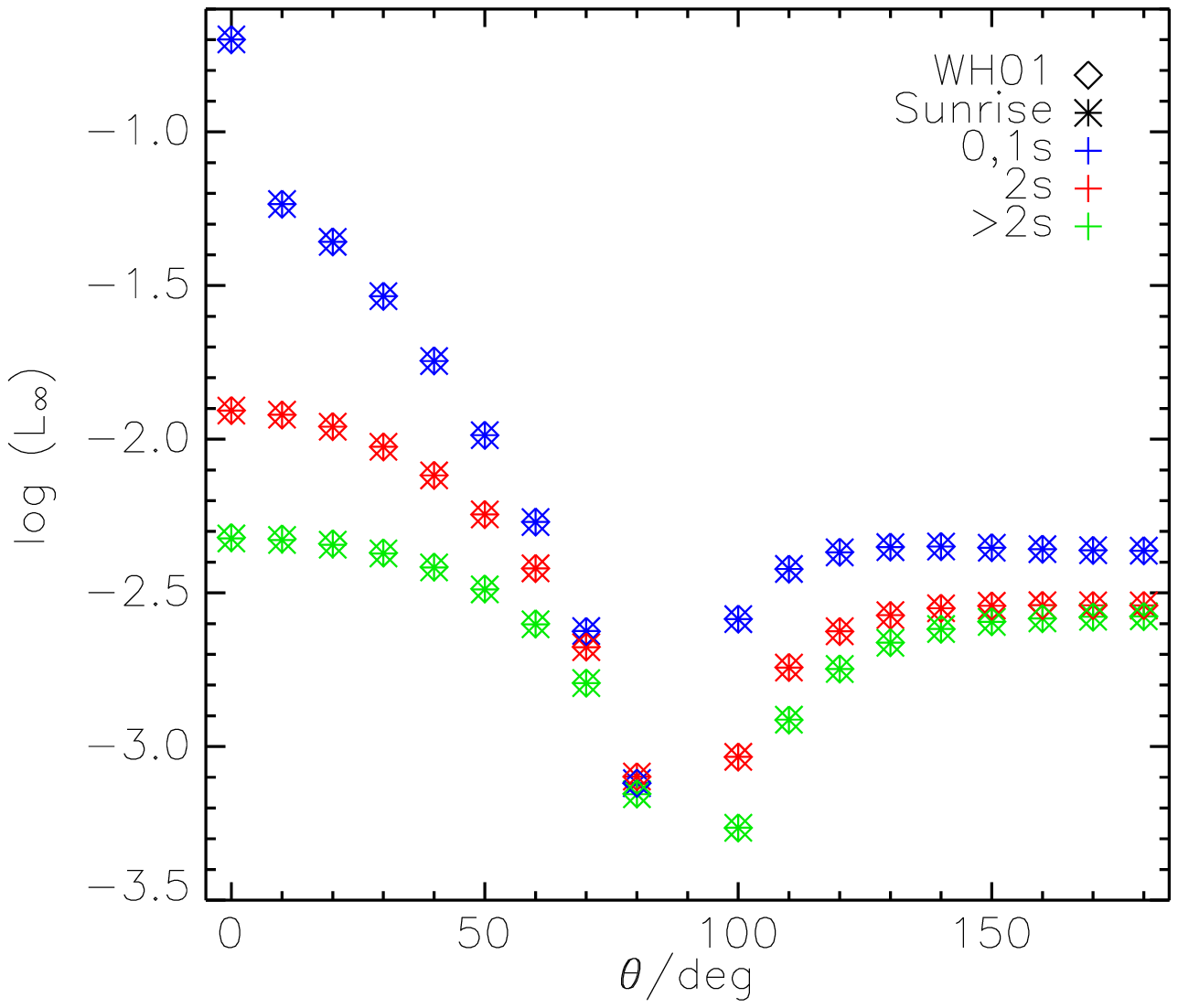} \end {center}  \caption{ \label{figure_wh01} Comparison with \citet{watsonhenney01}, showing the luminosity (or, more accurately, the radiative intensity) emerging in different directions from an infinite slab. On the left, the equivalent of their figure~1, where a point source is sitting on the surface of the slab, and on the right the equivalent of their figure~2, where a collimated beam is radiating into the surface. The WH01 results are shown as diamonds, the \mcrx\ results as asterisks. Direct and singly scattered light is in blue (WH01 separated this into direct and singly scattered light, but that is not easy to do by default in \mcrx), double-scattered light in red and triply scattered and above in green. The results agree to a remarkable degree.  } \end{figure*}

\citet{watsonhenney01} presented a problem geometry very similar to the
W77 study, and argued that this was a suitable test for
radiative-transfer codes in that it was simple but yet nontrivial.  In
this problem, an infinite plane-parallel slab of unit optical depth is
illuminated either by a point source on the surface of the slab or by a
collimated beam incident along the normal of the slab.  (While the
extent of the slab is described as infinite, this violates those
authors' own definition of the set of problems they are treating, in
which the opaque region is only of finite extent.  This restriction
applies also to \mcrx , so the slab was made finite but much larger
than its thickness.)  In contrast with W77, \citet{watsonhenney01} used
the ``radiative intensity'' instead of the surface brightness. The
radiative intensity is the total flux resulting from the source region,
at large distance, multiplied by the distance squared.  This quantity,
which is independent of distance, can be determined even in cases where
the source is unresolved.  Figure~\ref{figure_wh01} shows the radiative
intensity in different directions using the point source and collimated
beam.  The results have been split into three categories: direct plus
single-scattered light, doubly scattered light, and more-than-doubly
scattered light.  (\mcrx \ has no explicit facilities for selecting
light depending on its history, but by comparing runs with different
$I_{ \rm { min } }$, this can be extracted.)  The \mcrx \ results agree
remarkably well with the ones presented by \citet{watsonhenney01}.

\subsection{Comparison to Witt \& Gordon (1996)}
 \label{section-WG96}

\begin{figure} \begin {center} \includegraphics[width=0.99\columnwidth]{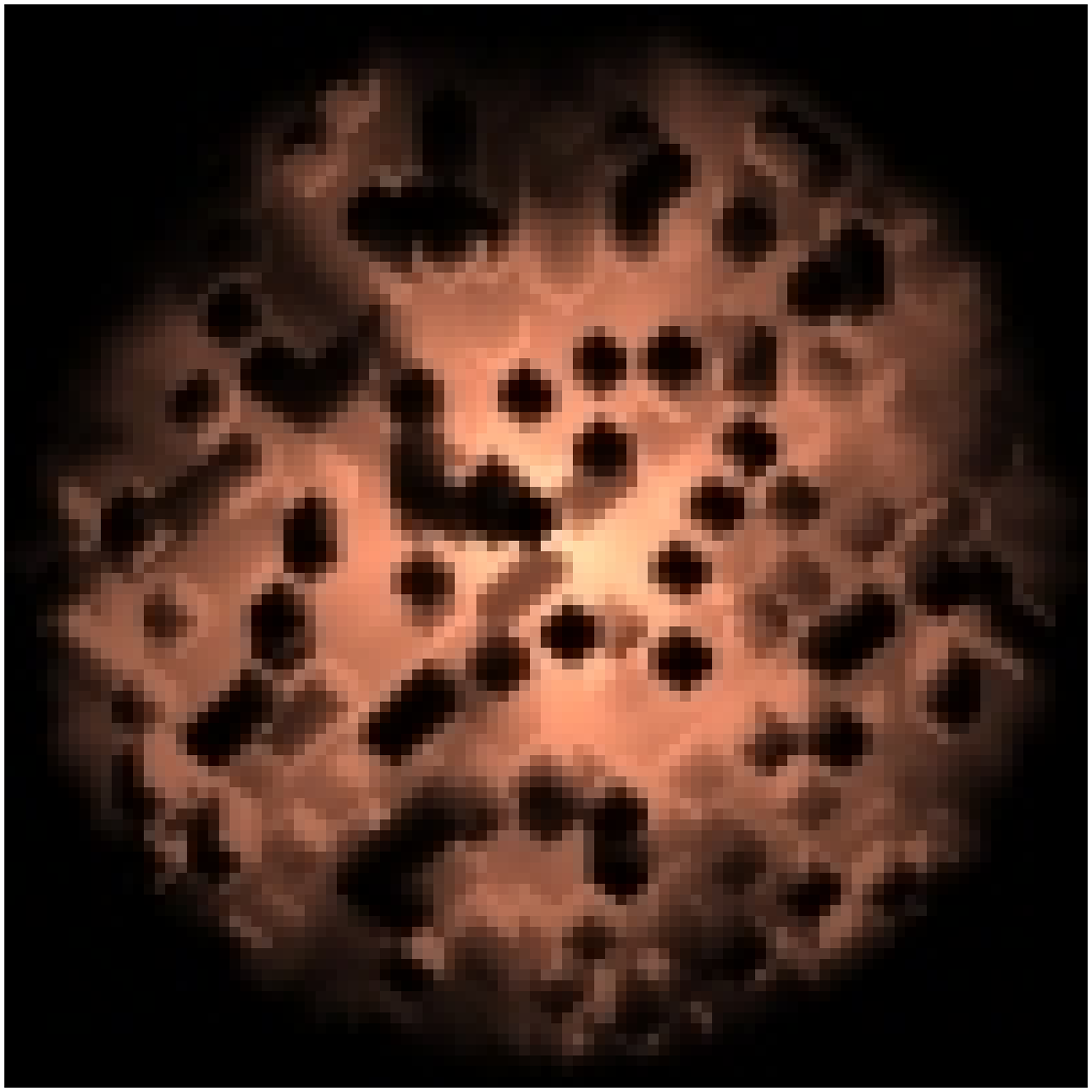} \end {center}  \caption[Radial surface brightness distribution]{ \label{wg96image} Image of the radiation emerging from a realization of the \citet{wittgordon96} random clumpy medium, with parameters $\tau_H=10$, $f\!f=0.10$, $k=0.01$, and $N=20$. Most radiation is emerging on paths through the low-density medium. The inner surfaces of the high-density clumps can be seen as bright edges due to the efficient scattering of light off of them, while dark shadows are cast towards the outside. } \end{figure}

As a more complicated test, the ``clumpy scattering environments'' of
\citet[][from now on WG96]{wittgordon96},
 was replicated and the results compared.  Briefly, WG96 investigated
the escape of radiation from a clumpy environment consisting of a
spherical region built up by smaller cubes of randomly assigned high-
or low-density material illuminated by a central point source. The
setup is described by 4 parameters: the filling factor, $f \! f$, the
fraction of cells which have high density; the density contrast, $k$,
the ratio of opacities between the low-density and high-density cells;
the grid resolution, $N$; and the optical depth of a homogeneous
distribution with the same amount of dust, $\tau_H$. WG96 also defined
the effective optical depth $\tau_{ \rm { eff } } = - \ln ( L_{ \rm {
dir } } / L_{ \rm { tot } } )$, where $L_{ \rm { dir } }$ is the direct
luminosity escaping through the distribution, without scattering, and
$L_{ \rm { tot } }$ is the source luminosity. An image of a realization
of these clumpy regions is shown in Figure~\ref{wg96image}.

This test is more difficult to perform conclusively, because the
problem geometry itself is random.  To get accurate results, not only
must the emerging luminosity be averaged over sufficiently many lines
of sight, but enough realizations of the clumpy medium must also be
used.  WG96 did not described the number of lines of sight or the
number of realizations of the medium used, but our results indicate
that at least 150 lines of sight and 60 realizations were necessary to
obtain reasonably converged results.  Even so, significant
uncertainties in the \mcrx results remain.  The uncertainties in the
results obtained by WG96 are unknown.

Due to the nature of the grid used by \mcrx , the problem cannot be
compared exactly; in WG96, the ray tracing was terminated when the rays
pass through a sphere inscribed within the grid, but this cannot be
done in \mcrx .  Instead, the adaptive grid was used to approximate a
spherical surface, leading to a slightly ``tiled'' surface, as opposed
to smooth, spherical one.  Four subdivisions were used, for which the
results were largely converged.  In any case, the uncertainty due to
this effect is dwarfed by the uncertainty due to the averaging over
lines of sight and realizations of the medium.

 \begin{figure*} \begin {center} \includegraphics[width= 0.49\textwidth]{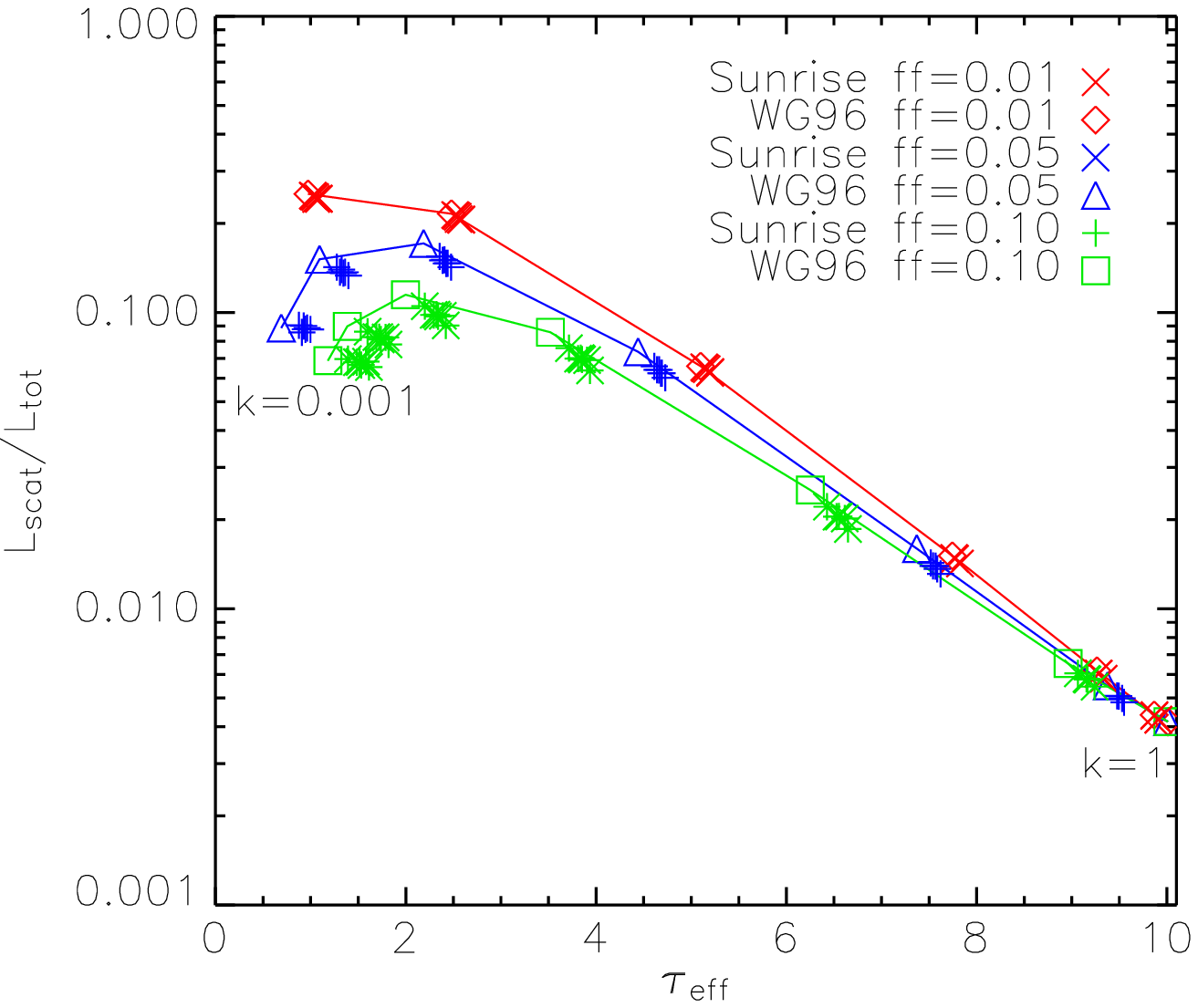} \includegraphics[width= 0.49\textwidth]{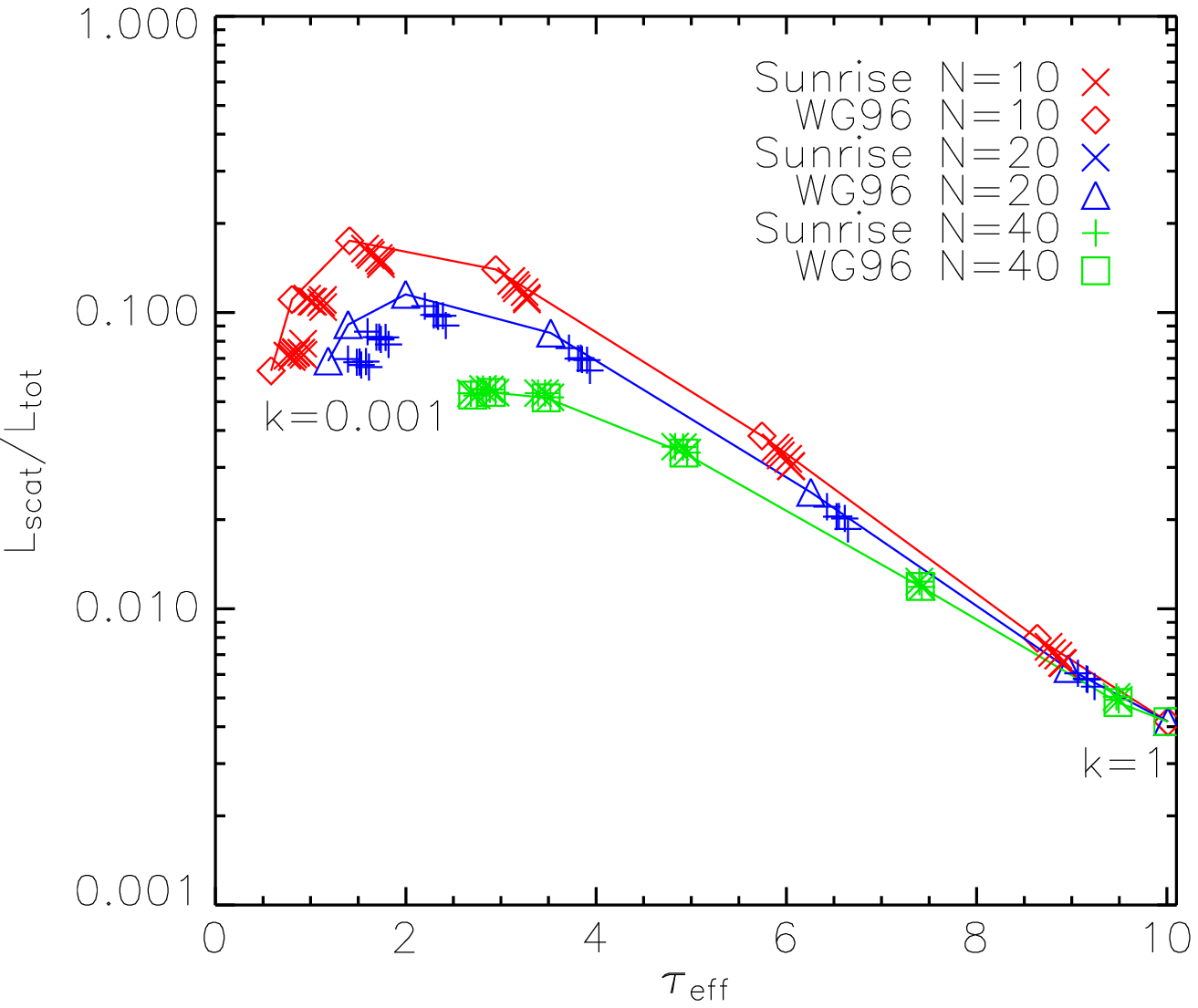} \end {center}  \caption{ \label{plot_wg96_fig89} Comparison with figures~8 -- 9 of \citet {wittgordon96}: scattered luminosity emerging from the clumpy system vs. effective optical depth, for a range of density ratios ($1\ge k_2/k_1\ge 0.001$) in steps of factors of $\sqrt{10}$, starting with $k=1$ at $\tau_{\mathrm{eff}}=10$. On the left, the equivalent of figure~8 in WG96, three different filling factors, $f\!f$, are shown in different colors, while on the right, the equivalent of figure~9 in WG96, the colors indicate different grid grid resolutions, $N$.  The WG96 points are connected by lines, while the corresponding \mcrx\ points, for a number of different random realizations, are found immediately to the right and below. In general, as the density contrast departs from unity, the effective optical depth is lowered, and the scattered luminosity increases. The \mcrx\ results agree with WG96 for the $k = 1 $ case, and for large $N$, but in other cases \mcrx\ predicts systematically higher $\tau_{\mathrm{eff}}$ and lower $L_\mathrm{scat}$.} \end{figure*} \begin{figure} \begin {center} \includegraphics[width=0.99\columnwidth]{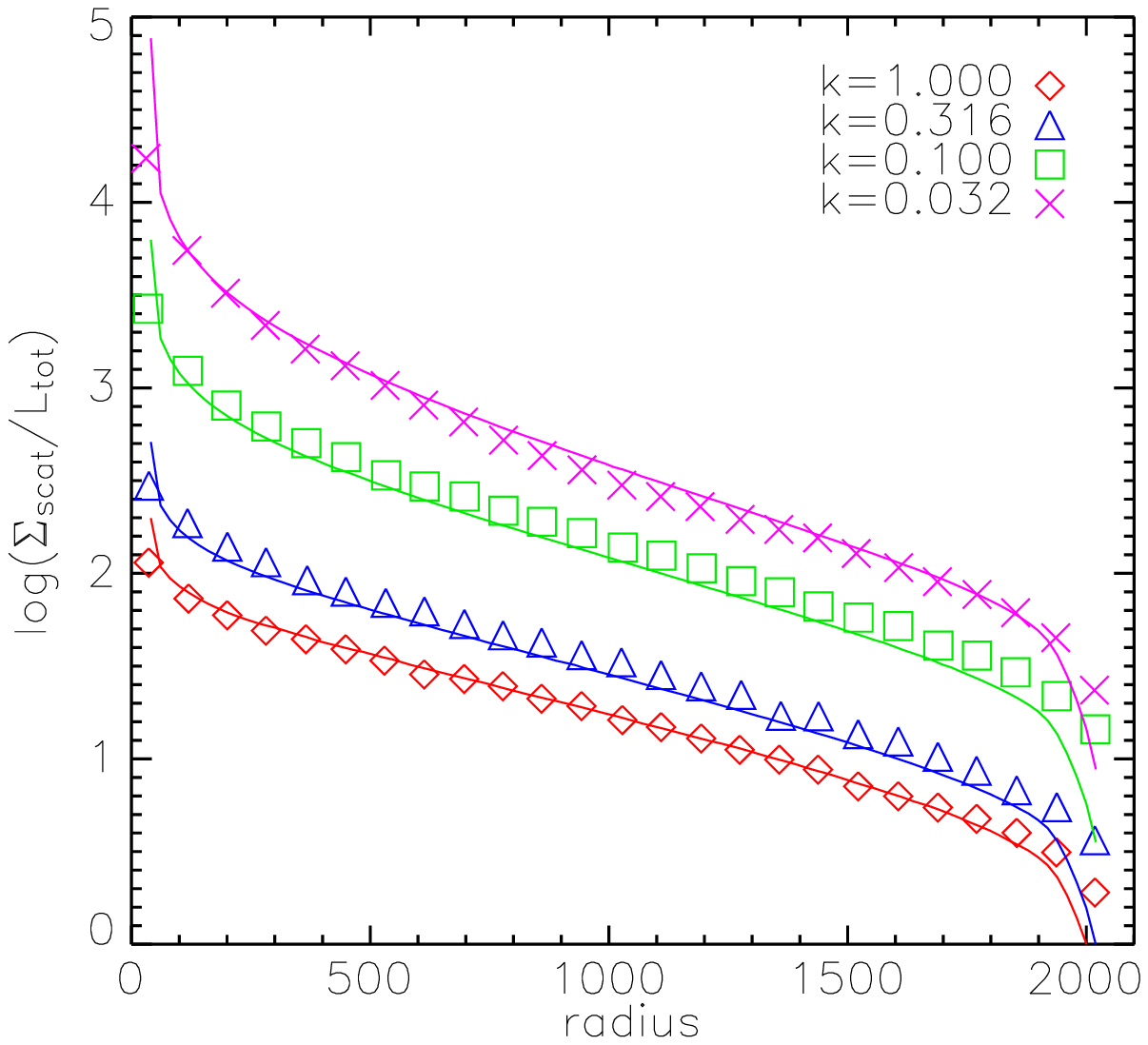}

\includegraphics[width=0.99\columnwidth]{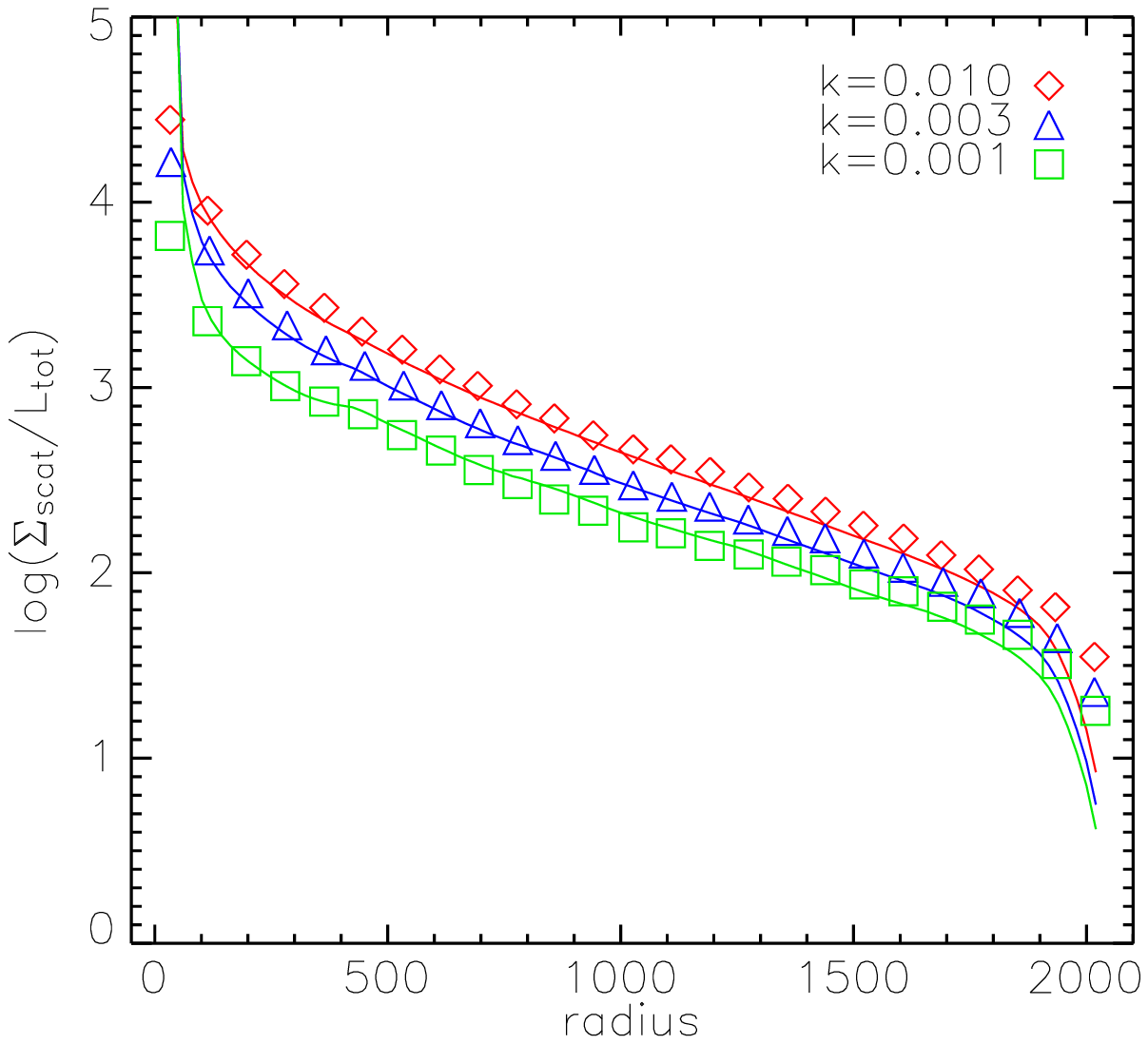} \end {center}  \caption[Radial surface brightness distribution]{ \label{plot_wg96_fig15} Radial surface brightness distributions of light for a range of density ratios ($1\ge k_2/k_1\ge 0.001$). This is the equivalent of figure~15 in \citet{wittgordon96}, but for clarity the different cases have been separated into two plots.  The \mcrx\ results are shown as lines, and the WG96 results as points.  The \mcrx\ results fall off more sharply towards the edge, but over a large range of radii the agreement is good.  The sharp rise in the \mcrx\ lines compared to the innermost WG96 point is because the direct light from the central point source is included in the \mcrx\ results, but has been removed in the WG96 data. } \end{figure}

Two tests are done.  The first is the equivalent of figures~8 and 9 in
WG96, where the ratio of the scattered luminosity, $L_{ \rm { scat }
}$, the luminosity escaping the system after scattering at least once,
to the total luminosity is plotted against $\tau_{ \rm { eff } }$ for
varying values of the density contrast $k$.  The results are shown in
Figure~\ref{plot_wg96_fig89}.  As the density contrast departs from
unity, the effective optical depth is lowered and the scattered
luminosity increases.  This is a natural consequence of clumping, as
low-density paths open up through the medium and allow radiation to
escape.  The \mcrx \ results agree with those of WG96 very well for the
homogenous case, but as the density contrast increases, the \mcrx \
points show systematically higher $\tau_{ \rm { eff } }$ than the WG96
ones. The discrepancy seems to decrease as the number of grid cells
increase; the $N = 40$ case agrees within the statistical uncertainty.

The statistical spread in the \mcrx \ points is strongly correlated. 
The main cause of this uncertainty is the random variation in the
number of high- vs. low-density cells in the grid. Realizations in
which the number of high-density cells is small will have smaller
$\tau_{ \rm { eff } }$ and larger $L_{ \rm { scat } }$.  Interestingly
enough, this variation is also generally in the direction towards the
WG96 results.  Experiments showed that artificially lowering $f \! f$
by about 0.015 from the $f \! f = 0.10$ case, while keeping the
opacities unchanged, will largely remove the discrepancy between the
\mcrx \ and WG96 points.  However, this does not work very well for $k
\la 0.01$, and is probably not the cause of the discrepancy, the origin
of which remains unknown.

The second test is the equivalent of figure~15 in WG96.  This compares
the radial surface brightness distribution of scattered light, averaged
over many viewing directions and realizations of the clumpy medium, for
the case $\tau_H = 10$, $f \! f = 0.10$ and $N = 10$. The results are
shown in Figure~\ref{plot_wg96_fig15}.  As the density contrast departs
from unity, the overall surface brightness of the scattered light
increases from the homogenous case, and the distribution becomes more
peaked towards the center.  For very small contrast ratios, when the
low-density cells are already optically thin, the surface brightness
again begins to decrease.

Because the WG96 figure is plotted with ``arbitrary units'' on the
x-axis, it has been assumed that the interval plotted is the full
extent of the object.  The units of their surface brightness are also
not specified, so the \mcrx \ results have been adjusted so that the $k
= 1$ case agrees with WG96.  The \mcrx \ lines agree well with the WG96
results over a large range of radii, but fall off more quickly towards
the edge.  The \mcrx \ results are also more sharply peaked compared to
the innermost bin of WG96, but this is because the direct light from
the point source in the center has not been removed.  There is also
agreement in some of the finer structure, for example a small bump in
the surface brightness at $r \approx 450$ for the $k = 0.001$ case,
corresponding to the radius of the innermost grid cell boundary.

As seen, there are some systematic differences between the WG96 results
and those determined using \mcrx .  However, the discrepancies are not
large, and both results share the same qualitative behavior. While it
would be desirable to pin down the cause of the discrepancies, we do
not believe that they are severe enough to cause alarm.

\subsection{Tests of the Polychromatic algorithm}
 \label{section_polychromatic_tests}

\begin{figure*} \begin {center} \includegraphics[width=0.33\textwidth]{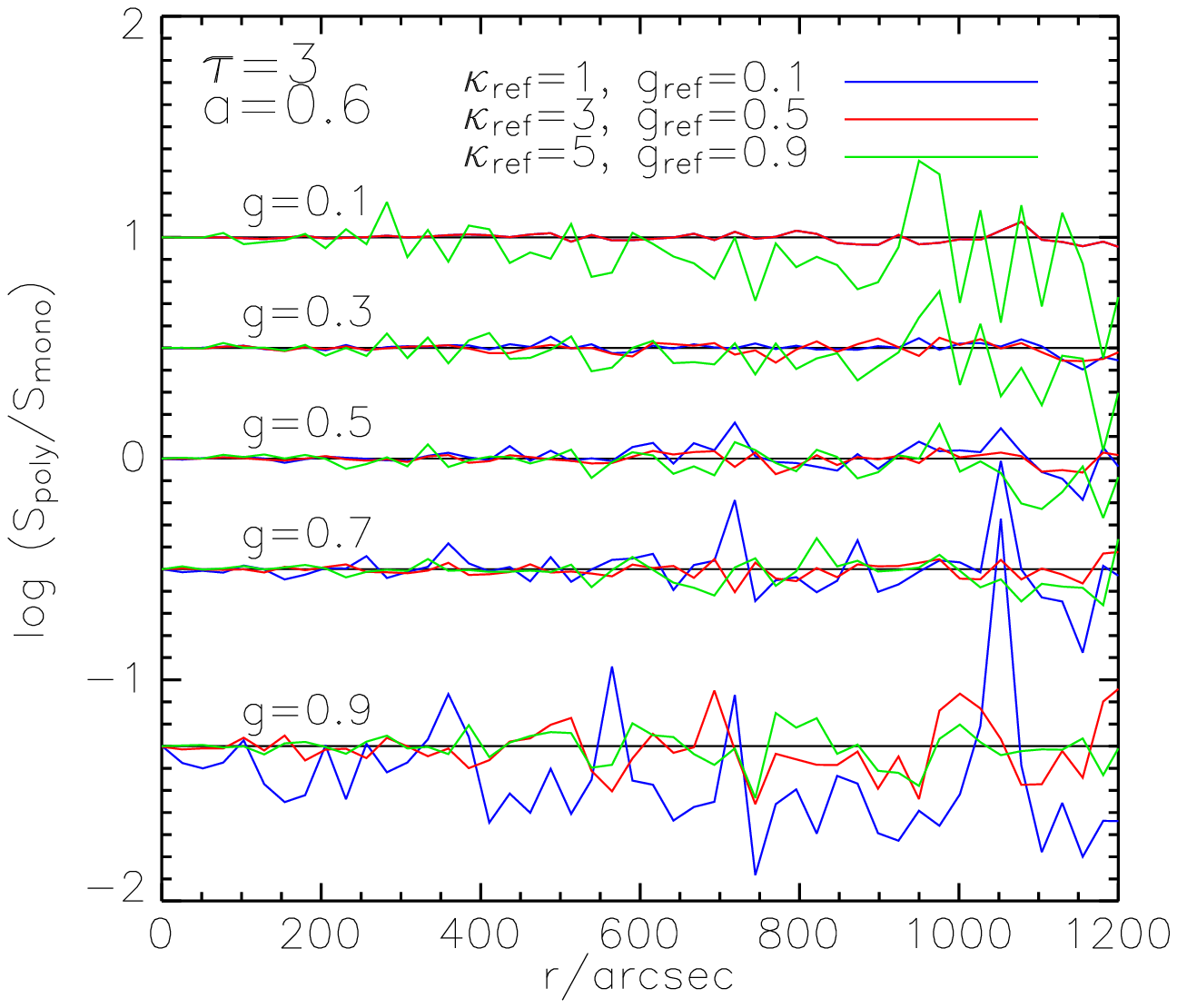} \includegraphics[width=0.33\textwidth]{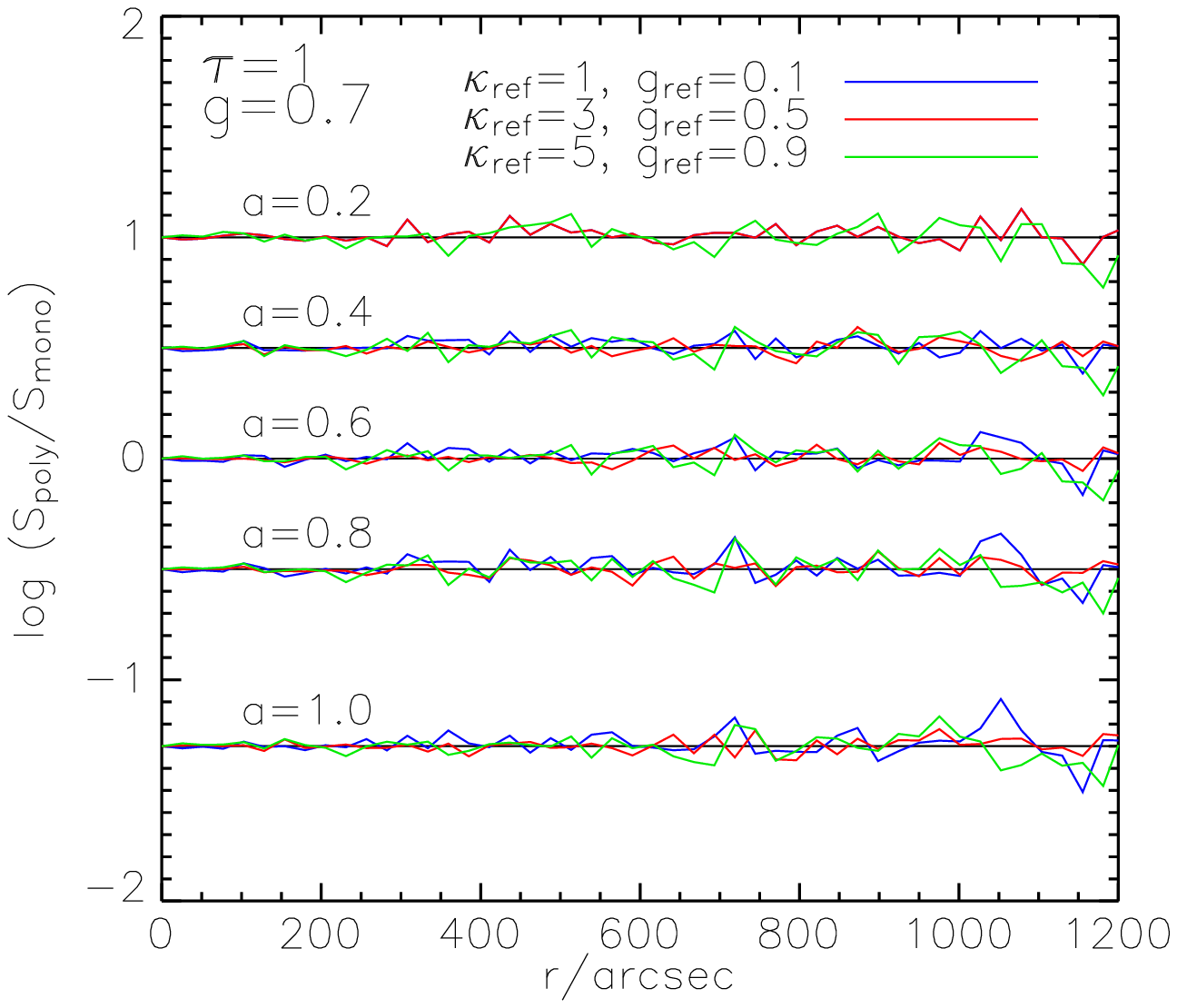} \includegraphics[width=0.33\textwidth]{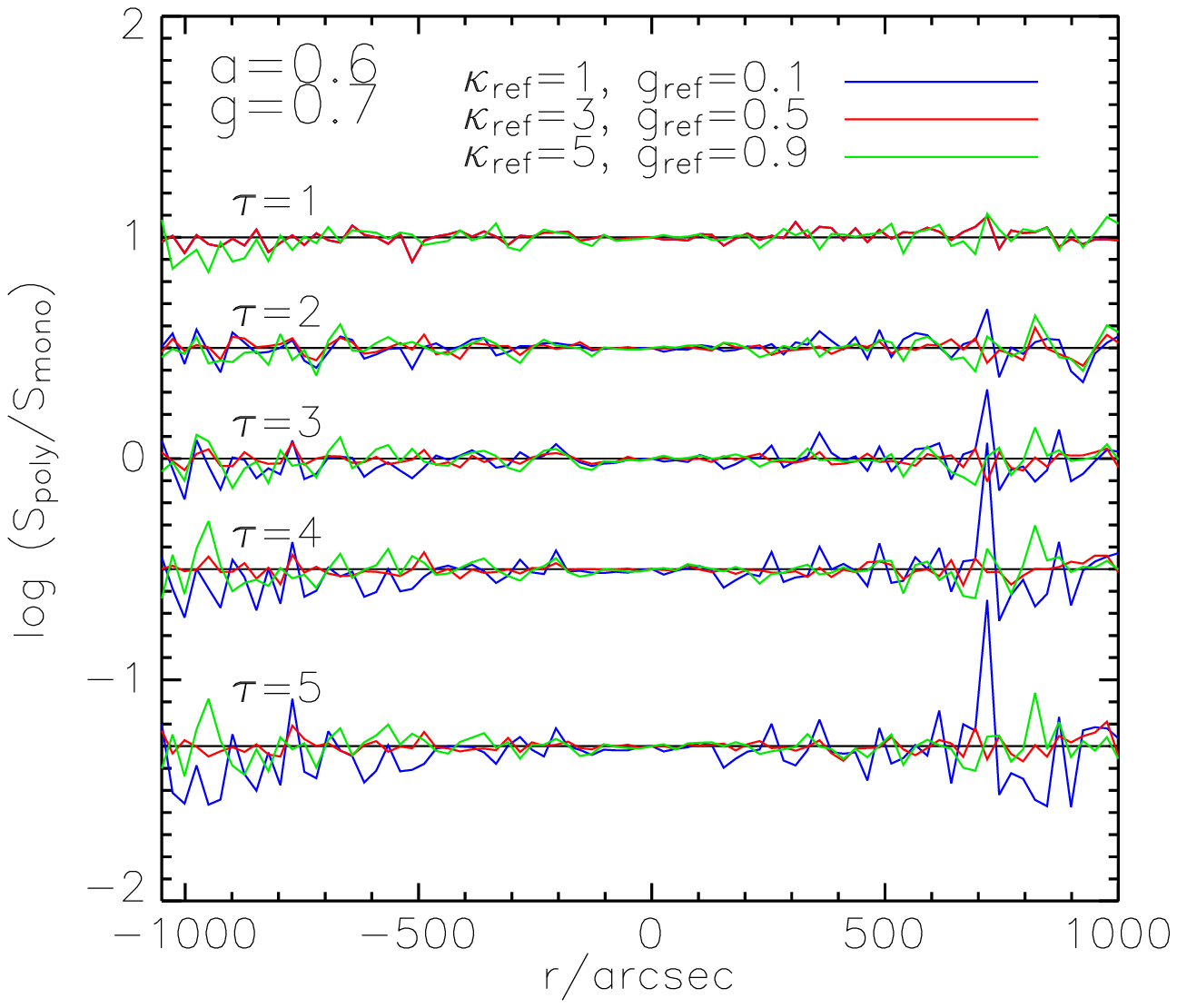} \end {center}  \caption{ \label{figure_polychromatic_w77} The polychromatic algorithm tested on the W77 problems.  The three plots show the deviation in surface brightness between the polychromatic results, where all 13 test cases were calculated simultaneously, and the monochromatic results presented in Figure~\ref{figure_w77}.  The different cases have been displaced from zero for clarity.  For each test case, results using three different reference opacities and scattering asymmetries are shown by different colored lines.  In general, the agreement is good.  There is significantly more noise in those cases where the reference values differ greatly from the case being tested, most notably the bottom blue line in the left plot where very forward-scattering dust ($g=0.9$) is being traced with practically isotropically scattering dust ($g_\mathrm {ref}=0.1$).  Increased noise is also evident in the top left green line ($g=0.1,\, g_\mathrm {ref}=0.9$) and the bottom right blue line ($\kappa=5,\, \kappa_\mathrm{ref}=1$).} \end{figure*} \begin{figure*} \begin {center} \includegraphics[width=0.99\textwidth]{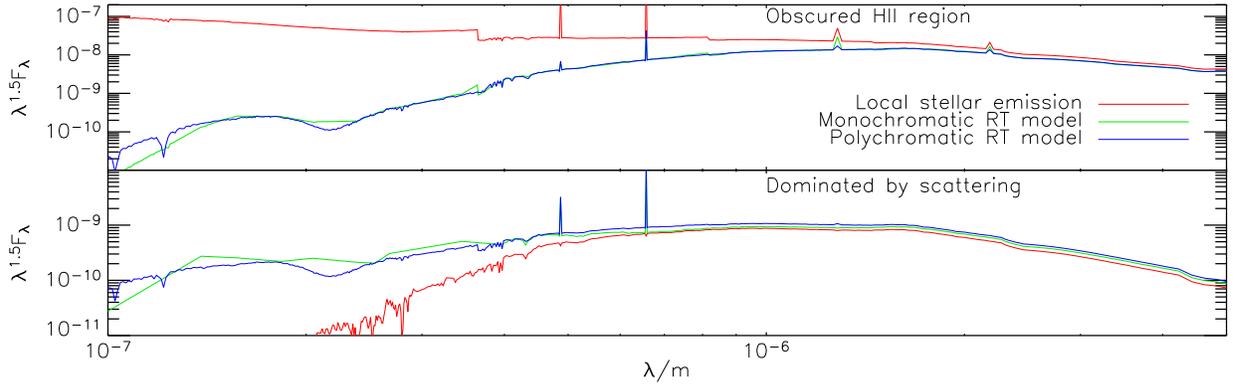} \end {center}  \caption{   \label{figure_polychromatic_SED}   Spectra from the galaxy merger simulations, exemplifying the effects of   the new polychromatic algorithm.  In both images, the red line shows   the intrinsic emission in the pixel, neglecting radiative-transfer   effects.  The green line shows the old algorithm, where the spectrum   is interpolated between 20 discrete wavelengths, and the blue line   is the result from the polychromatic radiative transfer.  The top   plot shows a pixel containing an obscured \hii\ region, the bottom   plot a pixel near an \hii\ region where the UV flux is dominated by   scattered light.  While the results agree well on the overall   spectral shape, the new method gives significantly more realistic   results especially for the small-scale spectral features, at a   fraction of the runtime.  Note in particular how the polychromatic   algorithm predicts the appearance of the Ly$\,\alpha$ absorption   line, the disappearance of the nebular Balmer continuum edge, and   the increased attenuation of the Paschen$\,\beta$ line at $1.3\um$   in the spectrum of the \hii\ region.  The polychromatic calculation,   including 500 wavelengths, used about 8 times the CPU time required   for one monochromatic calculation. With monochromatic ray tracing,   500 separate wavelengths would have to be used to predict the same   amount of detail, which would require a factor of 50 more CPU time. } \end{figure*}

The ``polychromatic'' algorithm has been tested with a preliminary
implementation, showing that it does indeed reproduce the results
obtained with the single-wavelength calculation, but it is not yet
included in the production version of \mcrx .

As a test, the W77 comparisons presented in Section~\ref{section-W77}
were redone, but now all different optical depths, albedos and
scattering asymmetries were calculated simultaneously. The results are
shown in Figure~\ref{figure_polychromatic_w77}. On the whole, the
agreement is excellent when the reference opacity and scattering
asymmetry is close to the case being calculated.  When the reference
parameters are very different from the case being calculated, the
results can indeed become very noisy.  This is most evident in the
cases where very forward-scattering dust ($g = 0.9$) is being
calculated using an isotropically-scattering reference case ($g_{ \rm {
ref } } = 0.1$, or vice versa.  In these cases, the weighting factor
can reach values of up to 200.

Very large weighting factors can also be obtained in very optically
thick cases where $\tau / \tau_{ \rm { ref } } < 1$. In fact, it is
clear from equation~\ref{equation-tau-bias} that in those cases
$w_\lambda$ will increase without bound while the total contribution
$w_\lambda exp ( - \tau_{ \rm { ref } } )$ from those cases remains
finite.  This suggests that convergence will be poor, and emphasizes
that a proper choice of reference parameters is crucial for the
accuracy of the method; the reference wavelength should be chosen such
that the range of weighting factors encountered in the problem is
minimized.

The polychromatic algorithm has also been tested on our galaxy merger
simulations.  This is a favorable situation, as direct light often
dominates. Examples of outputs are shown in
Figure~\ref{figure_polychromatic_SED}, which shows the spectra of two
different pixels in a galaxy merger simulations.

The dust opacity in a standard Milky-Way dust model varies by close to
three orders of magnitude from the far-UV to the near-IR.  The
disadvantages of the large weighting factors resulting from this large
range of opacity can be alleviated by stratifying the calculation into
several ranges of wavelengths, which will restrict the range of
opacities treated in the same random walk.  This is used in the next
section.

\subsection{The Pascucci et al. (2004) 2D benchmark}
 \label{section-P04}

\begin{figure*} \begin {center} \includegraphics[width= 0.49\textwidth]{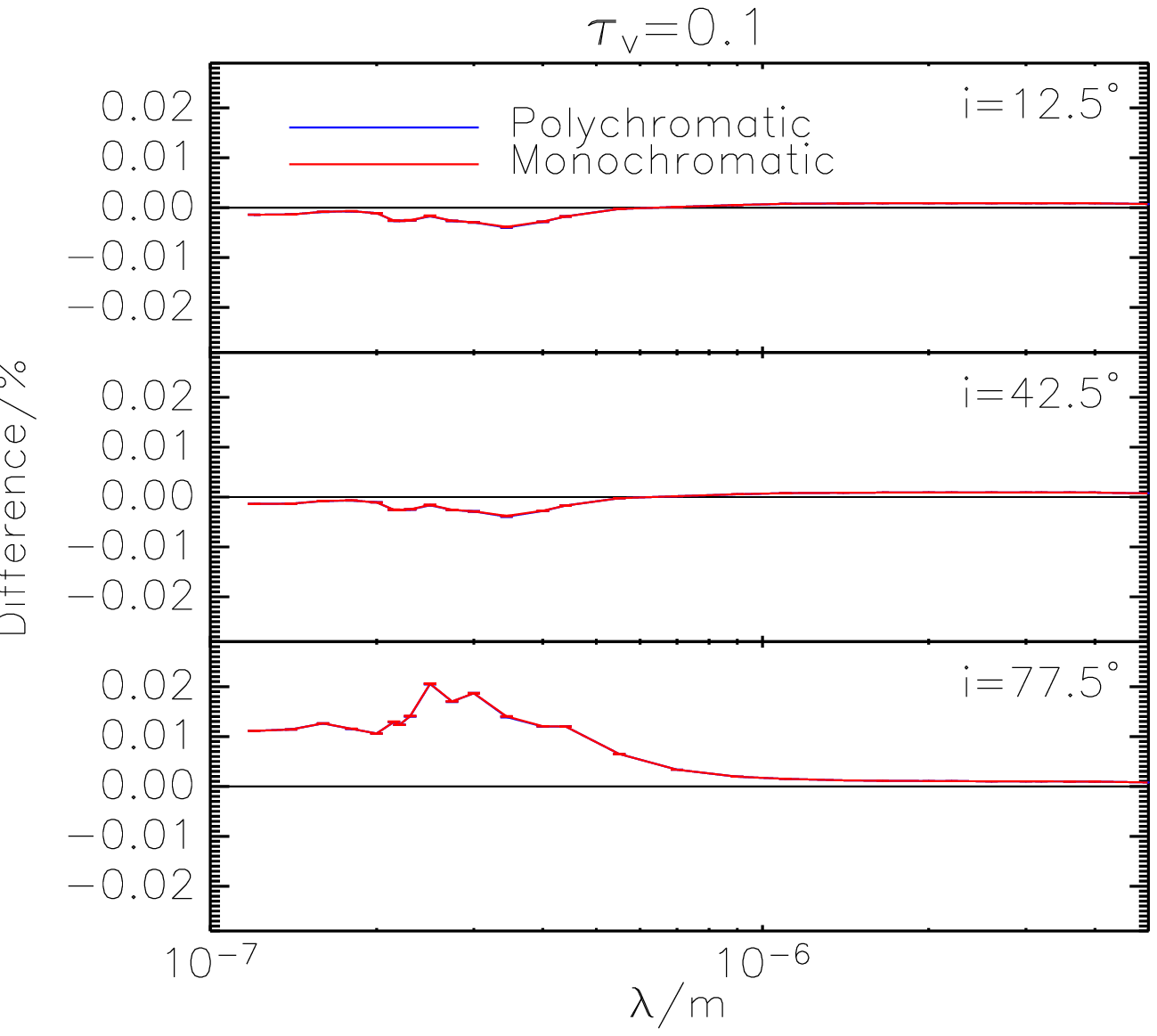} \includegraphics[width= 0.49\textwidth]{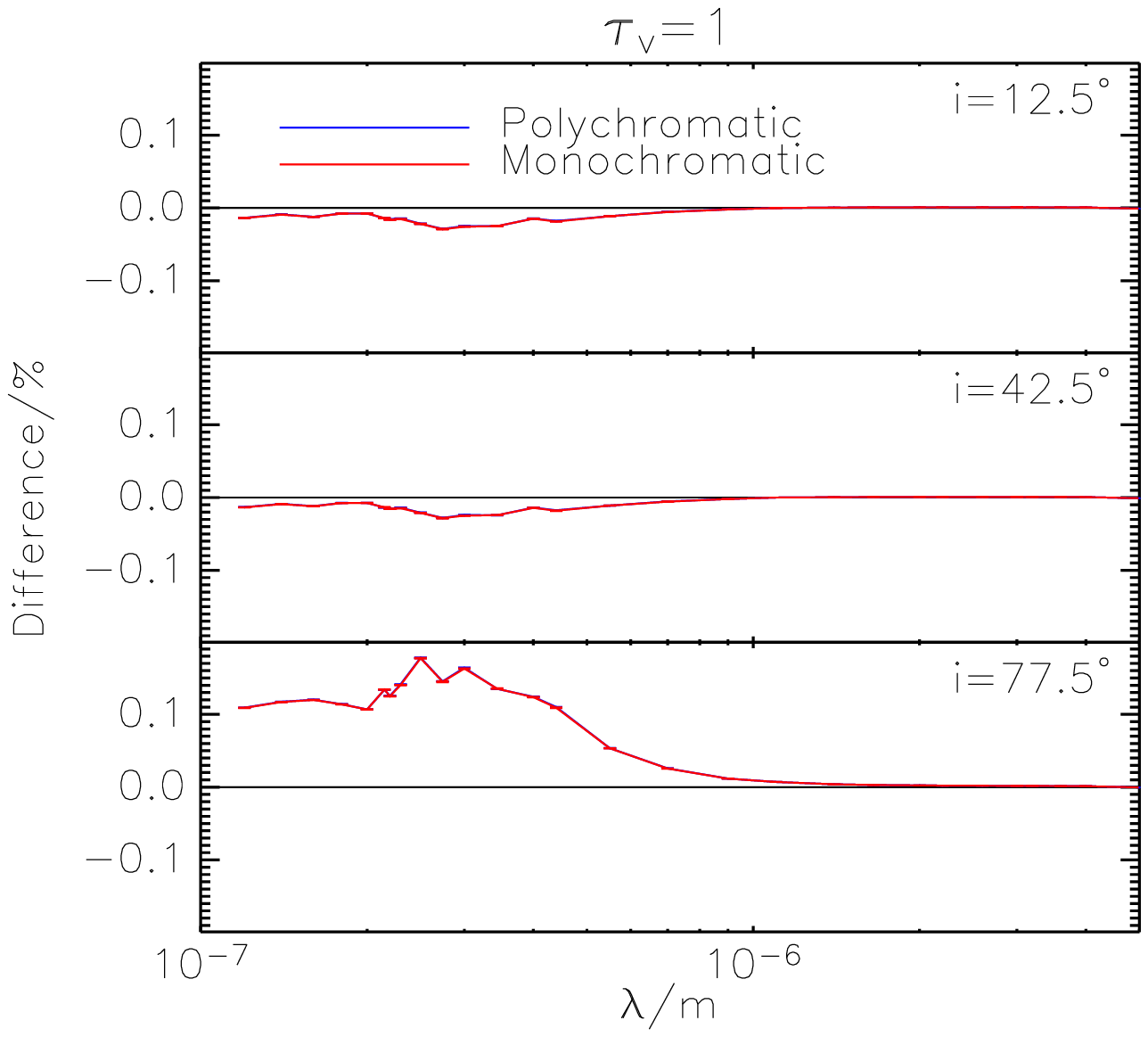}

\includegraphics[width= 0.49\textwidth]{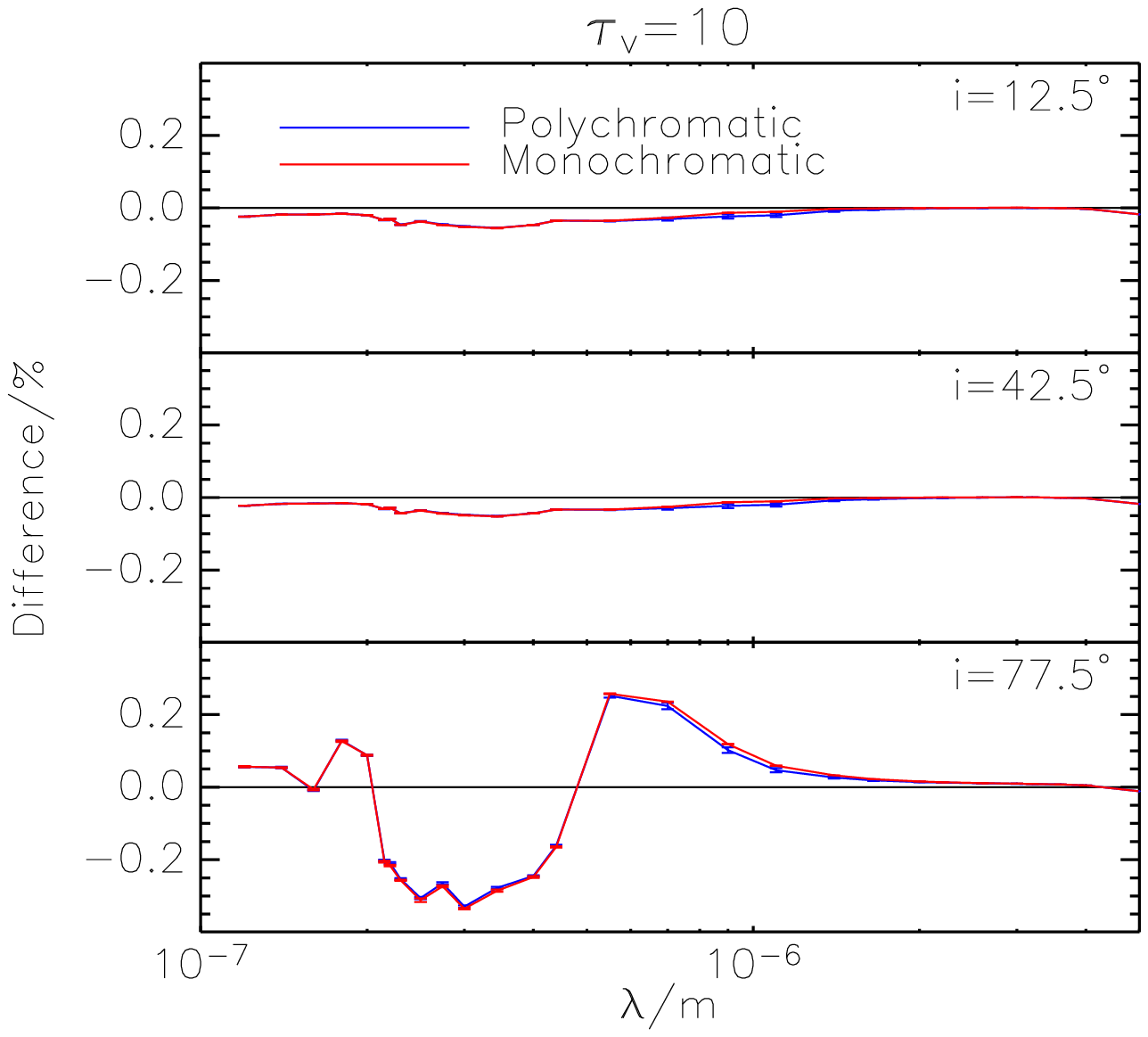} \includegraphics[width= 0.49\textwidth]{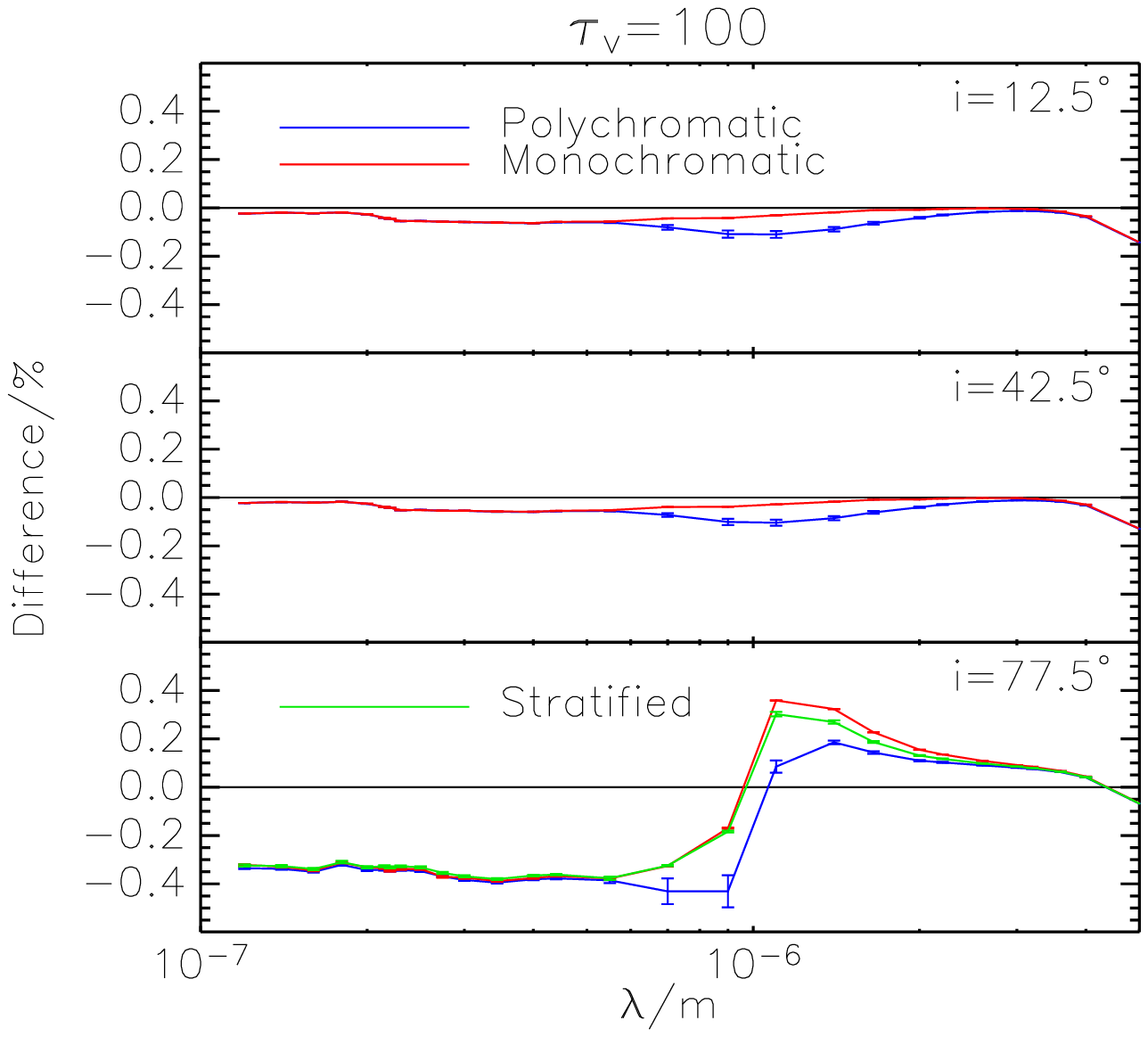} \end {center}  \caption{ \label{figure-P04-ratio} Difference in SED between the results obtained with \mcrx\ and the reference code {\sc radical} for the \citet {pascuccietal04} benchmark problem.  The monochromatic algorithm is shown in red, the polychromatic algorithm in blue.  For the $\tau_v=100$ edge-on case, the ``stratified'' polychromatic calculation (see text) is shown in green. The error bars (frequently too small to be visible) indicate the $1\sigma$ Monte-Carlo sampling uncertainty in runs with $10 ^ 6$ rays (per wavelength for the monochromatic algorithm).  The discrepancy reaches 40\% for the edge-on $\tau=100$ case. The results of the polychromatic algorithm are indistinguishable from those obtained with the monochromatic one except for the most optically thick case, where the results diverge around $1\um$. The stratified polychromatic calculation avoids this problem and agrees well with the monochromatic run. This figure is directly comparable to Figure~8 in \citet {pascuccietal04}.  } \end{figure*}

\citet[][from now on P04]{pascuccietal04}
aimed at establishing a standard benchmark radiative-transfer problem,
in the spirit of similar benchmark hydrodynamic tests, where a number
of codes are compared against each other in a more complicated problem
that lacks an analytical solution.  They designed a two-dimensional,
axisymmetric problem modeling a circumstellar disk and calculated the
dust temperature distribution and emerging SED from a number of
inclination angles.  The optical data for the dust grains and the
outputs from the 5 codes they used are available for download, which
makes a detailed comparison possible. Since \mcrx \ currently does not
have the capability to calculate temperature distributions, only the
SEDs at wavelengths shorter than $5 \um$, where stellar light
dominates, were compared. Comparisons were done both using the
monochromatic and polychromatic algorithms, which makes it possible to
evaluate the efficiency of the polychromatic method in a fairly
complicated problem.

The codes tested by P04 all used two-dimensional, axisymmetric grids
with variable grid spacings.  The orthogonal, explicitly
three-dimensional adaptive-mesh grid used by \mcrx \ is not ideally
suited for such a problem, as the number of grid cells necessary to
resolve the problem geometry well is much greater compared to a 2D
grid.  The P04 benchmark was also used by \citet{ercolanoetal05} to
test the MOCCASIN combined photoionization/dust radiative transfer
code, which uses a (uniform) Cartesian grid.

The \mcrx \ results are presented in Figure~\ref{figure-P04-ratio},
which is analogous to Figure~8 in P04. The differences between the
results using \mcrx \ and the RADICAL \citep{dullemondturolla00} code
are plotted as a function of wavelength for the four different optical
depths and three different inclination angles used by P04.  For small
optical depths the results agree very well, which is not surprising
since the SED is close to the intrinsic blackbody SED of the central
source.  For larger optical depths, and especially for the edge-on
configurations, the discrepancies reach $\pm 40 \%$. This is larger
than the internal differences between the codes used by P04, and is
probably due to the use of an orthogonal three-dimensional grid.  This
hypothesis is supported by the fact that the results were converging
(slowly) towards the P04 results as the grid resolution was increased. 
The results plotted used a grid with $1.7 \times 10^6$ cells and a
minimum cell size of $0.02 \times 0.02 \times 0.002 \> \mbox { AU }$. 
Grids with up to $4.3 \times 10^6$ cells have been tried, and improved
the agreement of around $5 \%$ in the most optically thick case. The
agreement between the P04 benchmark and the \mcrx \ results is still
significantly better than those presented by \citet{ercolanoetal05} for
the MOCCASIN code, using a uniform Cartesian grid, which disagree with
the P04 results by up to a factor of 20.  It is unclear what grid
resolution was used by \citet{ercolanoetal05}.

\begin{figure} \begin {center} \includegraphics[width= 0.49\textwidth]{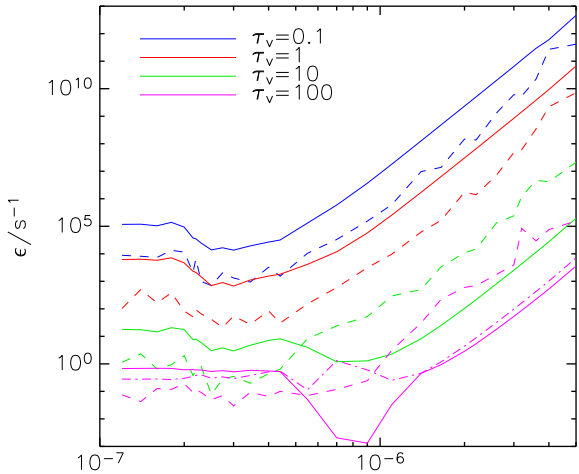} \end {center}  \caption{ \label{figure-P04-efficiency} The efficiencies of the polychromatic and monochromatic methods in the \citet{pascuccietal04} benchmark problem. The polychromatic algorithm is shown as solid lines, the monochromatic as dashed lines, while the color indicates the optical depth of the problem.  For the $\tau_v=100$ case, the efficiency of the ``stratified'' polychromatic calculation is shown as a dot-dashed line.  For lower optical depths, the efficiency of the polychromatic algorithm exceeds that of the monochromatic one for all wavelengths.  For $\tau_v=10$, the \emph{minimum} efficiency, indicating the error of the most poorly constrained wavelength, is still higher for the polychromatic algorithm.  For the highest optical depth, the polychromatic calculation suffers from the large range of optical depths for different wavelengths, and shows very low efficiencies around $0.8\um$.  The stratified calculation, where wavelengths longer than $0.5\um$ are traced separately, avoids this problem and keeps the minimum efficiency significantly higher than the monochromatic algorithm. } \end{figure}

The benchmark was calculated using both the monochromatic and
polychromatic methods. For the monochromatic calculations, $10^6$ rays
were used for each of the 28 wavelengths used by P04 blueward of $5
\um$, while the polychromatic calculations used $10^6$ polychromatic
rays and a reference wavelength of $0.4 \um$.  Except for the $\tau_v =
100$ case, the results are indistinguishable in
Figure~\ref{figure-P04-ratio}.  As the optical depth increases, the
polychromatic calculation becomes increasingly affected by the large
range of optical depths encountered at different wavelengths.  The
reference wavelength used was $0.4 \um$, where both opacity and albedo
are high, and as a consequence accuracy suffers at wavelengths between
$0.7 \um$ and $1.5 \um$, where the opacity drops but the albedo is
still fairly high.  This problem was solved by ``stratifying'' the
calculation into two wavelength ranges.  The reference wavelength was
kept at $0.4 \um$ for wavelengths shorter than $0.55 \um$, while longer
wavelengths were traced separately, using a reference wavelength of
$0.7 \um$.  This two-step calculation agrees very well with the
monochromatic results.

To quantify the relative efficiencies of the two methods, the
``efficiency'', $\epsilon$, defined as 
\begin{equation}
\epsilon = \frac { F_\lambda^2 } { T \sigma_{ F_\lambda }^2 } ,
\end{equation}
 where $T$ is the CPU time required to complete the calculation and
$\sigma_{ F_\lambda }$ is the Monte Carlo sampling uncertainty in the
SED, is calculated. The efficiency defined in this way is insensitive
to the number of rays traced and quantifies the (inverse of the) CPU
time necessary to produce results of unit relative accuracy.
Figure~\ref{figure-P04-efficiency} plots the efficiencies of the
monochromatic and polychromatic methods as a function of wavelength for
the edge-on cases in Figure~\ref{figure-P04-ratio}.  No effort was made
to optimize the relative number of rays at different wavelengths in the
monochromatic calculation.

For $\tau_v < 10$, the efficiency of the polychromatic method exceeds
that of the monochromatic one for all wavelengths.  As mentioned above,
the polychromatic algorithm begins to suffer at higher optical depths. 
For $\tau_v = 10$, the efficiency of the monochromatic algorithm
overtakes the polychromatic one at wavelengths longer than $0.5 \um$. 
However, the \emph{minimum} efficiency, which corresponds to the
wavelength with the largest error in the SED, is still higher for the
polychromatic algorithm.

For the very optically thick $\tau_v = 100$ case, the efficiency of the
polychromatic algorithm plummets for the same wavelength range where
the errors were large in Figure~\ref{figure-P04-ratio}, reaching a
minimum value much lower than the monochromatic calculation. The
stratified polychromatic calculation, however, maintains an efficiency
greater than the monochromatic calculation out to $1 \um$ and has a
minimum efficiency significantly larger than the monochromatic
results.

It is important to remember that this example used a comparably low
number of 28 wavelengths.  While the run time for the monochromatic
method will increase linearly as the wavelength resolution is improved,
the polychromatic method will scale much better (at least until the run
time becomes dominated by the vector calculations as opposed to the ray
tracing).  It is clear that the particular strength of the
polychromatic algorithm is in processing densely sampled wavelengths,
not in covering a large range in wavelength.

\section{Code Implementation Details }
 \label{implementation_section}

\mcrx \ is written in C++, with a highly modular code structure. It is
easy to add new types of emission, sources of scattering, etc., as
these are isolated from the core radiative-transfer engine. Efficient
vector calculations are provided by the Blitz++ library \citep{blitz}.

\mcrx \ uses the implementation of the Mersenne Twister MT19937
random-number algorithm \citep{mersenne_twister} provided by the
Blitz++ library \citep{blitz}.

\subsection{Grid Traversal}

\begin{figure} \begin {center} \includegraphics[width= 0.99\columnwidth]{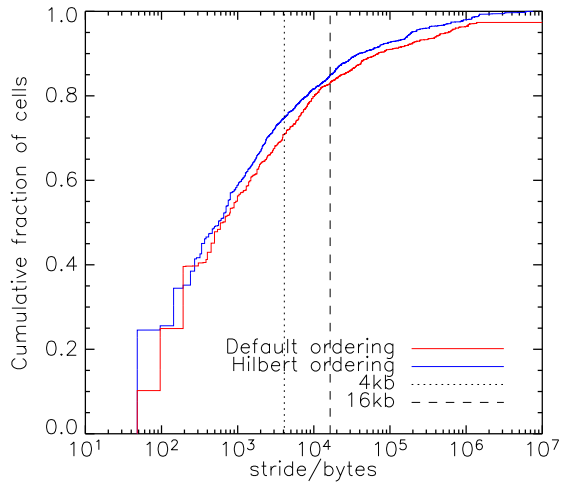} \end {center}  \caption[Memory access pattern during grid traversal]{ \label{plot_strides} The memory access pattern in a test run of \mcrx.  The graph shows the cumulative distribution of strides (differences in memory address) when accessing consecutive grid cells during a run consisting of about $10 ^ 5 $ grid-cell-to-grid-cell steps.  The minimum stride is 48 bytes, the size of the grid-cell data structure, and this is the source of the discretization of the small strides.  The vertical lines indicate 4096 and 16384 bytes, the size of the TLB entries on the POWER3/Opteron and Altix 300 processors, respectively. The Hilbert ordering results in noticeably smaller strides than the default depth-first ordering, but not enough to outweigh its increased complexity.} \end{figure}

The most time-consuming part of the calculation is the grid traversal,
walking the ray from grid cell to grid cell.  Because the rays can
traverse the grid in any direction, the memory access pattern of the
code is hard to predict, and often not optimal.  To minimize CPU cache
and TLB (Translation Lookaside Buffer) misses, it is important to
access memory in as localized a fashion as possible.  This is easy in
applications like matrix multiplications, where it is up to the
programmer to decide in which order to access memory, but hard in the
application considered here.  The solution is to order the grid cells
in memory in such a way that locality is preserved; grid cells that are
close in space should also be stored close in memory.  Since this is a
mapping from 3D to 1D, it cannot be accomplished perfectly, and various
approaches exist.  The strategy used by \mcrx \ is to simply store the
subgrids consecutively in memory in the order they are encountered when
traversing the tree of refined cells in a depth-first fashion. This
actually results in a fairly compact memory access pattern.  To explore
the potential for improvements, storing the grid cells according to a
Hilbert curve was also tried, using the algorithm described in
\citet{bartholdigoldsman01}.  The Hilbert curve is an example of what
is known as ``space-filling curves'', a one-dimensional curve which
fills three-dimensional space, and has been argued as being nearly
optimal for these problems \citep{niedermeier96}.  The result of this
comparison is shown in Figure~\ref{plot_strides}.  While the Hilbert
ordering did have a more compact access pattern, the differences were
comparatively small and did not outweigh the accompanied increased
complexity.  Thus, the default memory ordering was retained.

In addition, finding neighboring cells in the grid is quite expensive,
as the tree has to be traversed up the hierarchy and then down again,
accessing many higher-level grid cells in the process.  Because of
this, \mcrx \ utilizes a caching scheme whereby each cell saves
pointers to its neighboring cells once they have been determined.  This
speeds up subsequent neighbor searches considerably.

\subsection{Runtime Requirements}
 \label{section-run-time}

The most obvious scaling is that the runtime of the ray tracing
(excluding initialization, I/O, etc.)  obviously scales linearly with
the number of rays traced.  The time required per ray is affected by
many factors, including the problem geometry, the importance of
scattering, and the terminating intensity $I_{ \rm { min } }$.  The
time is dominated by the tracing of the ray from one grid cell to the
other, which takes approximately constant time.  Emission and
interaction events are much more rare and do not constitute the
dominant computational load.  From this it would be expected that the
time required to trace one ray would scale with the number of grid
cells, which is the case.  The scaling is roughly $t_r \sim N_c^{ 1 / 3
}$, which would also be the naive expectation since this is how the
number of grid cells traversed by a ray should scale with the total
number of cells.

Furthermore, increasing the number of cameras increases the
computational load, since part of the ray propagation consists of
calculating $\tau^{ \rm { obs } }$, which means traversing the cells
from the point of emission or interaction   to each of the observers. 
The time taken to complete one Monte-Carlo ray can thus be split into
two components such that $t_r = t_0 + N_c t_c$, where $t_0$ is the time
taken to complete the ray tracing in the absence of observers, and
$t_c$ the time taken to traced the optical depth to an observer.  This
relation is indeed obeyed by the code, and the two components $t_0$ and
$t_c$ are such that, in typical cases of simulated galaxies, the time
proportional to $N_c$ becomes dominant for $N_c \ga 3$.

\subsection{Parallelization}

Any high-performance numerical code relies on parallel execution to
reach high performance, so this capability was a basic requirement for
the development of \mcrx . The Monte-Carlo method is easy to
parallelize; as long as the entire grid fits in memory, each ray is
independent of others.  \mcrx \ uses multithreaded (shared-memory)
parallelism during the grid creation and ray-tracing stages.  As every
ray is independent, communication requirements are minimal resulting in
essentially perfect scaling as long as the memory bus is not
saturated.

Completing the full calculation for one simulation snapshot takes 6-12
CPU hours on a contemporary dual-CPU 2.2GHz Opteron system, for typical
conditions of 11 camera positions, 21 wavelengths and $10^6$ rays per
wavelength.  The total amount of CPU time consumed by the
radiative-transfer calculations is roughly equal to the time used by
the hydrodynamic galaxy merger simulations.

On the NASA Altix 3000 system Columbia, a cluster of machines each
consisting of 256 two-processor nodes, where the full, system-wide
memory is visible as a single address space using high-speed
interconnects, \mcrx \ can be used on up to 16 processors with a
penalty of only 23 percent compared to running on one node.

Adding the ability to run using distributed memory would vastly
increase the size of the problems that could be treated.  However, it
would also make the parallelization much more complicated, and as the
current ability is sufficient for our computational needs, such an
upgrade is not currently planned.

\subsection{Distribution}

As a service to the community, \mcrx \ is available under the terms of
the GNU General Public License
\citep{gpl}\footnote{The \mcrx\ source code, documentation, and   example outputs are available on the \mcrx\ web site at   \url{http://sunrise.familjenjonsson.org}}.
Other users are encouraged to use \mcrx \ for their radiative-transfer
applications, and to add enhancements that increase its capabilities.

\section{Future Improvements}
 \label{future_section}

There are numerous improvements that can be made to the
radiative-transfer code.  The polychromatic algorithm is currently
being implemented in the production version of \mcrx , and schemes for
minimizing the impact of large weighting factors will be explored. One
possibility is to split rays with large weights and scatter them in
different directions \citep{mcprimer}, which will increase the sampling
in the heavily weighted part of phase space.  Such a scheme could be
combined with ``Russian roulette'' into a scheme where the code
attempts to keep the weights of all rays within some specified range.

The biasing of path lengths done in the polychromatic algorithm opens
up other possibilities; it is well known that the Monte Carlo method
has problems treating very large optical depths, for example in the
case of a cloud heated by external radiation studied by
\citet{juvela05}, because few rays make it into the opaque regions. 
\citet{juvela05} explored the effects of biasing the distributions of
the external photons and the scattering direction.  Another possibility
would be to bias the path lengths between scatterings to larger values,
which would allow better sampling of the rare photons which penetrates
unusually deep into the cloud.

These possibilities suggest that it would be advantageous to make it
possible to not only use arbitrary scattering phase functions or
emission distributions in \mcrx , but also to provide an infrastructure
for customized ``biasing plug-ins'' which can be selected by the user
depending on the problem at hand.

Another obvious improvement is to include a self-consistent calculation
of the dust emission, so predictions of infrared observations of
galaxies with e.g. the Spitzer Space Telescope can be made. These
calculations can be done to varying degrees of sophistication, from
calculating equilibrium temperatures of single grain species to full
stochastic temperature distributions of very small grains and PAH
emission, but it becomes necessary to include the heating of dust
grains due to the emission from other grains.  This introduces a
coupling between the local emissivity and radiation intensity, which
makes it necessary to integrate until the dust temperature distribution
converges.

An interesting alternative, applicable in situations where the opacity
is not a function of temperature, is the method of
\citet{bjorkmanwood01}. In this method, the dust temperatures are
updated whenever an absorption event happens and the energy is
reemitted with an SED equal to the \emph{difference} between the
current emission SED and the one before the absorption event happened. 
This ensures that the net radiation emitted by the dust is appropriate
for the temperature of the dust grains, and as more rays are traced the
dust temperature distribution will relax toward the equilibrium value
without iteration.  \citet{baesetal05} studied this method and
concluded that while it produces the correct frequency distribution in
the case of grains in thermal equilibrium, the method will not work for
stochastically heated grains because the probability distribution
required for the reemitted radiation will become negative.  As negative
probabilities are unphysical, the method will fail.  The polychromatic
algorithm suggests a way around this obstacle: There is no obvious
reason why the weights of the rays in the polychromatic algorithm can't
be negative.  The problem is not fundamentally that negative
probabilities are necessary, but that the grain has emitted too much
energy in a certain part of the spectrum and this must be corrected
for.  This can be accomplished with negative weights for certain
wavelengths.  Since it will only serve to remove energy which was
previously emitted, the radiation field should still converge towards
the true value, only it will not converge from below as in the original
\citet{bjorkmanwood01} formulation. This possibility will be explored
in \mcrx \ in the future.

Another important improvement for correctly predicting the SED of
galaxies is a more detailed modeling of star-forming regions, which are
not resolved in our galaxy simulations.  Studies have shown that extra
attenuation of young stellar populations are necessary for fitting dust
attenuations in galaxies \citep{silvaetal98, charlotfall00,
tuffsetal04}.  While the adaptive grid could be used to resolve these
regions, the computational cost would be prohibitive.  A better way
would be to use sub-resolution models of the emission from star-forming
regions
\citep[e.g.][]{grovesetal04mappings}
and feed this emission into the \mcrx \ grid.

\section{Conclusion}

This paper has described \mcrx , a new Monte-Carlo code for calculating
the radiative transfer of light through a scattering and absorbing
medium.  \mcrx \ builds on previous advanced Monte-Carlo codes
\citep{kurosawaetal04, baesetal03, gordonetal01, bianchietal00, lucy99,
wolfetal99}, and adds a polychromatic algorithm, where all wavelengths
are traced simultaneously, and efficient parallel computation in a
flexible, modular package, making calculations with a spatial dynamical
range of more than $10^4$ feasible.  Images at any wavelength from
far-UV to near-IR from an arbitrary number of directions are generated,
as well as the radiation intensity and dust luminosity as a function of
position in the object. In addition, \mcrx \  includes a framework for
calculating the effects of dust in hydrodynamic simulations, and the
code is freely available to the community.

Accurate radiative-transfer calculations with realistic geometries are
crucial in tying theory to observations in any situation where dust
significantly affects the radiation emerging from an object. This
applies to such diverse a family of objects as galaxies, star-forming
regions, AGN, and protostellar disks.  The outputs from the
radiative-transfer calculations generate ``simulated observations'' of
the object, so comparisons with observations can be done using
observational instead of theoretical quantities. This approach requires
specifying many free parameters, but in reality only serves to make
this dependence explicit, since if the comparisons are done with
theoretical quantities, the same parameters normally need to be
specified to convert observed quantities to intrinsic ones.  In many
cases, such as the complicated radiative-transfer situations that can
be solved by \mcrx , it is not even possible to invert the
observations, and making observational predictions from a theoretical
model is then the only avenue possible. Another advantage of comparing
observational quantities is that it is usually easier to mimic
selection biases and instrumental effects in simulations than to infer
their effects on theoretical quantities.

\mcrx \ is currently being used to investigate dust attenuation in
simulations of isolated spiral galaxies (Rocha Gaso et al., in
preparation), to generate a library of images of simulated galaxy
mergers which can be directly compared to Hubble Space Telescope
observations, and to quantify the timescales over which mergers result
in disturbed morphologies.

\medskip The author wishes to thank T. J. Cox and Joel Primack for
their help with the infrastructure to process the galaxy merger
simulations, and them as well as Sandy Faber, Greg Novak, Mike Kuhlen
and Brent Groves for stimulating discussions about the radiative
transfer problem. Thanks also go to A. Witt and A. Watson for their
assistance in pinning down the discrepancy with the W77 results, to the
referee, J. Yates, for comments which improved the paper, and to I.
Pascucci for help in setting up the P04 benchmark. The author is also
grateful for the hospitality of the Max-Planck Institute for
Astrophysics in Garching, where part of this paper was written, and for
support from a UC/LLNL cooperative grant from IGPP to Wil van Breugel.
This work was supported by program number HST-AR-10678.01-A, provided
by NASA through a grant from the Space Telescope Science Institute,
which is operated by the Association of Universities for Research in
Astronomy, Incorporated, under NASA contract NAS5-26555.  This research
used computational resources of the National Energy Research Scientific
Computing Center (NERSC), which is supported by the Office of Science
of the U.S.  Department of Energy, and the NASA Advanced Supercomputing
Division (NAS).

\bibliographystyle{../../../bib/mn2e} 
\bibliography{../../../bib/patriks} \label{lastpage} \end {document}